\documentclass[12pt]{article}
\usepackage{xr}
\usepackage{graphicx,subfigure,amsthm,amsmath,latexsym,amssymb,asa,enumitem}
\usepackage{float,epsfig,multirow,rotating,times,verbatim,wrapfig}
\usepackage[margin=0.9in]{geometry} 
\usepackage{mdwlist} 
\usepackage[affil-it]{authblk} 
\RequirePackage{setspace}
\RequirePackage{caption}
\usepackage[small,compact]{titlesec}
\titlespacing{\section}{0pt}{*1}{*1}
\titlespacing{\subsection}{0pt}{*1}{*1}
\titlespacing{\subsubsection}{0pt}{*1}{*1}
\titlespacing{\paragraph}{0pt}{*1}{*1}
\titlespacing{\subparagraph}{0pt}{*1}{*1}

\setlist{nolistsep}

\newenvironment{packed_enum}{
\begin{enumerate}[leftmargin=1.2em]
  \setlength{\itemsep}{1pt}
  \setlength{\parskip}{0pt}
  \setlength{\parsep}{0pt}
}{\end{enumerate}}

\usepackage[T1]{fontenc}
\usepackage{lmodern}
\usepackage[T1]{fontenc}
\usepackage{tgtermes} 
\RequirePackage{natbib}
\newcommand{\R}{I\!\!R}

\newcommand{\bepsilon}{\boldsymbol{\epsilon}}

\newcommand{\btau}{\boldsymbol{\tau}}

\newcommand{\bI}{\boldsymbol{I}}

\newcommand{\bY}{\boldsymbol{Y}}
\newcommand{\bZ}{\boldsymbol{Z}}

\newcommand{\bzero}{\boldsymbol{0}}

\newcommand{\bone}{\boldsymbol{1}}

\newcommand{\mI}{\mathcal I}
\newcommand{\mE}{\mathcal E}
\newcommand{\mV}{\mathcal V}
\newcommand{\E}{\mbox{I\!E}}

\newcommand{\apprasym}{
 \mathrel{\ooalign{$\sim$\cr\kern+1.25pt\large $\colon$}}}

\newcommand{\blinded}{0}

\begin{document}

\if0\blinded
{
\markboth{Ye et al{\it et al.}}{Automated Flaw Detection in NDE
  Images}
}
\fi
\if1\blinded
{
\markboth{}{Automated Flaw Detection in NDE Images}
}
\fi

\title{\vspace{-0.8in} 
A Statistical Framework for Improved Automatic Flaw Detection
  in Nondestructive Evaluation Images} 
\if0\blinded
{
\author{
  \vspace{-0.1in}
  {\bf Ye Tian}\\
  \vspace{-0.2in}
  Department of Statistics\\
    Iowa State University\\
    Ames, IA 50011\\
    (tianye1984@gmail.com)\\
\and
    \vspace{-0.2in}
    {\bf Ranjan Maitra}\\
    \vspace{-0.2in}
    Department of Statistics\\
    Iowa State University\\
    Ames, IA 50011\\
    (maitra@iastate.edu)
    \and
    \vspace{-0.2in}
           {\bf William Q. Meeker}\\
           \vspace{-0.2in}
           Department of Statistics and the Center for Nondestructive
           Evaluation\\
           Iowa State University\\
           Ames, IA 50011\\
           (wqmeeker@iastate.edu)
           \and
           \vspace{-0.2in}
                  {\bf Stephen D. Holland}\\
                  \vspace{-0.2in}
                  Department of Aerospace Engineering and Center for Nondestructive Evaluation\\
                  Iowa State University\\
                  Ames, IA 50011\\
                  (sdh4@iastate.edu)
}
\date{\vspace{-0.5in}}
}
\fi
\if1\blinded
    {
      \date{\vspace{-1.2in}}
      \renewcommand{\baselinestretch}{1.4}\normalsize
    }
\fi
\maketitle
\renewcommand{\baselinestretch}{1}\normalsize
\begin{abstract}
Nondestructive evaluation (NDE) techniques are widely used to detect
flaws in critical components of systems like aircraft engines, nuclear
power plants and oil pipelines in order to prevent catastrophic
events. Many modern NDE systems generate image data. In some
applications an experienced inspector performs the tedious task of
visually examining every image to provide accurate conclusions about
the existence of flaws. This approach is labor-intensive and can cause 
misses due to operator ennui. Automated evaluation methods 
seek to eliminate human-factors variability and improve throughput. Simple
methods based on peak amplitude in an image are sometimes employed and
a trained-operator-controlled refinement that uses a dynamic threshold
based on signal-to-noise ratio (SNR) has also been implemented. We
develop an automated and optimized  detection procedure that mimics
these operations. The primary goal of our
methodology is to reduce the number of images requiring expert visual
evaluation by filtering out images that are 
overwhelmingly definitive on the existence or absence of a flaw. We
use an appropriate model for the observed values of the SNR-detection
criterion to estimate the probability of detection. Our methodology
outperforms current methods in terms of its ability to detect flaws.
\end{abstract}
\vspace{-0.1in}
\renewcommand{\baselinestretch}{1.35}\normalsize
\section{Introduction}
Nondestructive evaluation (NDE)
methods~\citep{brayandstanley96,shull02,heller12} are used to examine
and characterize materials  and detect flaws in components without
causing them irreversible damage. Different physical
principles~\citep{rummel83,silketal87,brayandmcbride92} guide
different NDE techniques, providing methods based on
radiography~\citep{halmshaw82,halmshaw91},
ultrasound~\citep{krautkramerandkrautkramer90},
eddy-currents~\citep{kahnetal77,collinsetal85,yangetal10},
radiology~\citep{martzetal02}, active
thermography~\citep{spicerandosiander02}, acoustic 
emissions~\citep{prosser02}, magnetic particles~\citep{lindgrenetal02},
liquid penetrants~\citep{halmshaw91} and other
techniques~\citep{shull02,heller12}. One 
of the primary purposes of NDE is to detect flaws in critical system
components. Examples include fatigue cracks in aircraft engine turbine
disks  or blades, material anomalies in billets or forging materials
that can be detected  during manufacturing processes. Flaw detection
is important in almost all cases  but especially when there is a risk
of such flaws causing serious or disastrous  damage to systems
(e.g., aircraft or bridges).  Characterization of flaws after
detection is also important in the application of nondestructive
inspection, because it helps maintenance engineers obtain
knowledge about flaw types, shape, size, location, and
orientation, and use this information to decide on
whether a specific part should continue in service or be immediately
repaired or replaced.

There is substantial and long-standing interest in the development of
statistical methodology and algorithms for the analysis of
NDE data~\citep[{\em e.g.}, see][]{berensandhovey81,berensandhovey82,berensandhovey83,
berensandhovey84,grayandthompson86,annisanderland89,burkeletal96,hoveyandberens88,perdijon88a,perdijon88b,perdijon89,nealandspeckman93,howardandgilmore94,sweeting95,spencerandschurman95,olinandmeeker96,howardetal98,aokiandsuga99,zakietal01,legendreetal01,meyerandcandy02,zavaljevskietal05,dogandzicandzhang07,hasanzadehetal08,liandmeeker09,lietal10,gaoandmeeker12,ngetal13}. Much
of the NDE literature has revolved around the issues of noise
reduction and comcomitant increased signal-to-noise ratio (SNR)  and
development of methodologies~({\em e.g.}, development of wavelet methods,
expectation-maximization algorithm-type methods) for better estimating
the extent and probability of detection (POD) of a flaw in different 
techniques. Some other attempts have been in the area of automated
inspections. A comprehensive review of statistical issues and
development in NDE is provided in the discussion paper of
\citet{olinandmeeker96}. Another review of available
methodology and techniques in this research area is provided in
\citet{milhdbk09}. 

With the rapid advances in technology, there has also been a quantum
jump in the development of NDE techniques. More modern
and automated methods of data acquisition have also resulted in the
capability of obtaining high-quality  
image-based data. Such datasets have obvious advantages in
that images are  natural objects for a technician to examine and use
in order to make decisions. It is often much  easier and more
straightforward to detect the existence of a flaw or to assess flaw
characteristics ({\em e.g.}, size and orientation) from visual
inspection of an image than from a series of numbers provided by
traditional methods. Image data, however, also provide challenges in 
terms of interpretation and detection as illustrated next in the
context of vibrothermography which also forms the showcase
application of this paper.

\subsection{Analysis of Vibrothermography Image Data}
\label{vbexample}
Vibrothermography -- also called sonic infrared, thermoacoustics or
thermosonics -- is a modern NDE imaging
technique~\citep{maldague01,hennekeandjones79,reifsnideretal80,holland07}   
for detecting cracks or flaws in industrial, dental and aerospace 
applications. The imaging modality works on the principle that a sonic
or ultrasonic energy pulse when applied to a unit  causes
it to vibrate. As a result, it is expected that the faces of a crack will
rub against each other, resulting in an increase in temperature in
that region. An infrared camera captures this increased temperature and 
produces a sequence of images of the temperature intensities over a short period
of time starting just before the pulse of energy is applied and ending around
the time that generated heat has dissipated. A sequence of images
records the temperature changes over time. The primary objective of 
this technology is to detect flaws in the material with high
precision. If the crack is larger than a certain threshold,
the part needs to be 
repaired or replaced. 
Another goal is to predict the progression of the flaw whose sizes
are below a certain threshold and therefore not cause for immediate
concern, but important enough to suggest a purposeful schedule for
future inspections. 

\begin{figure*}[h]
\vspace{-0.05in}
\begin{center}
\mbox{
\subfigure[Strong signal]{
 \includegraphics[width=0.31\textwidth]{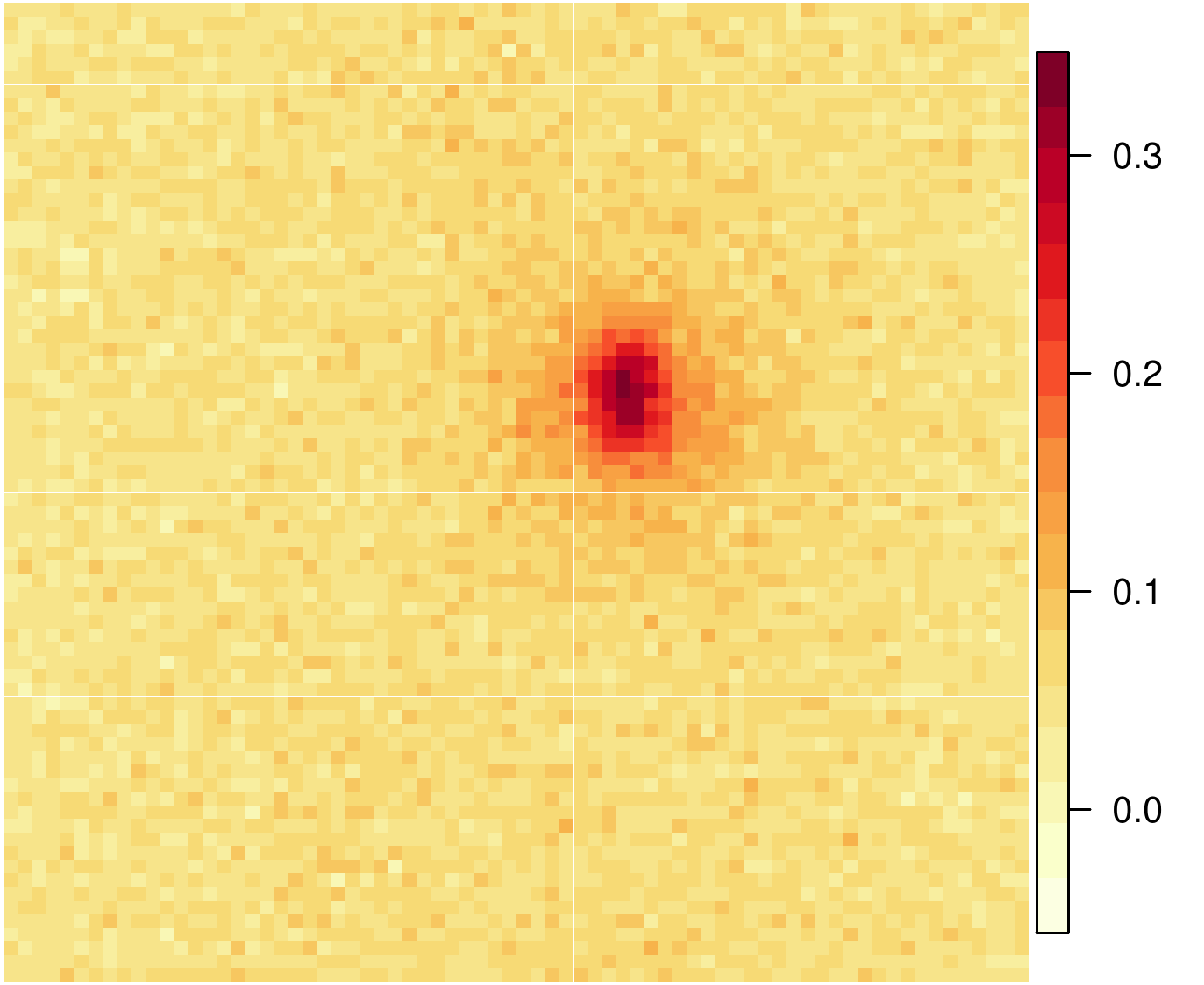}}
\subfigure[Weak signal]{
 \includegraphics[width=0.31\textwidth]{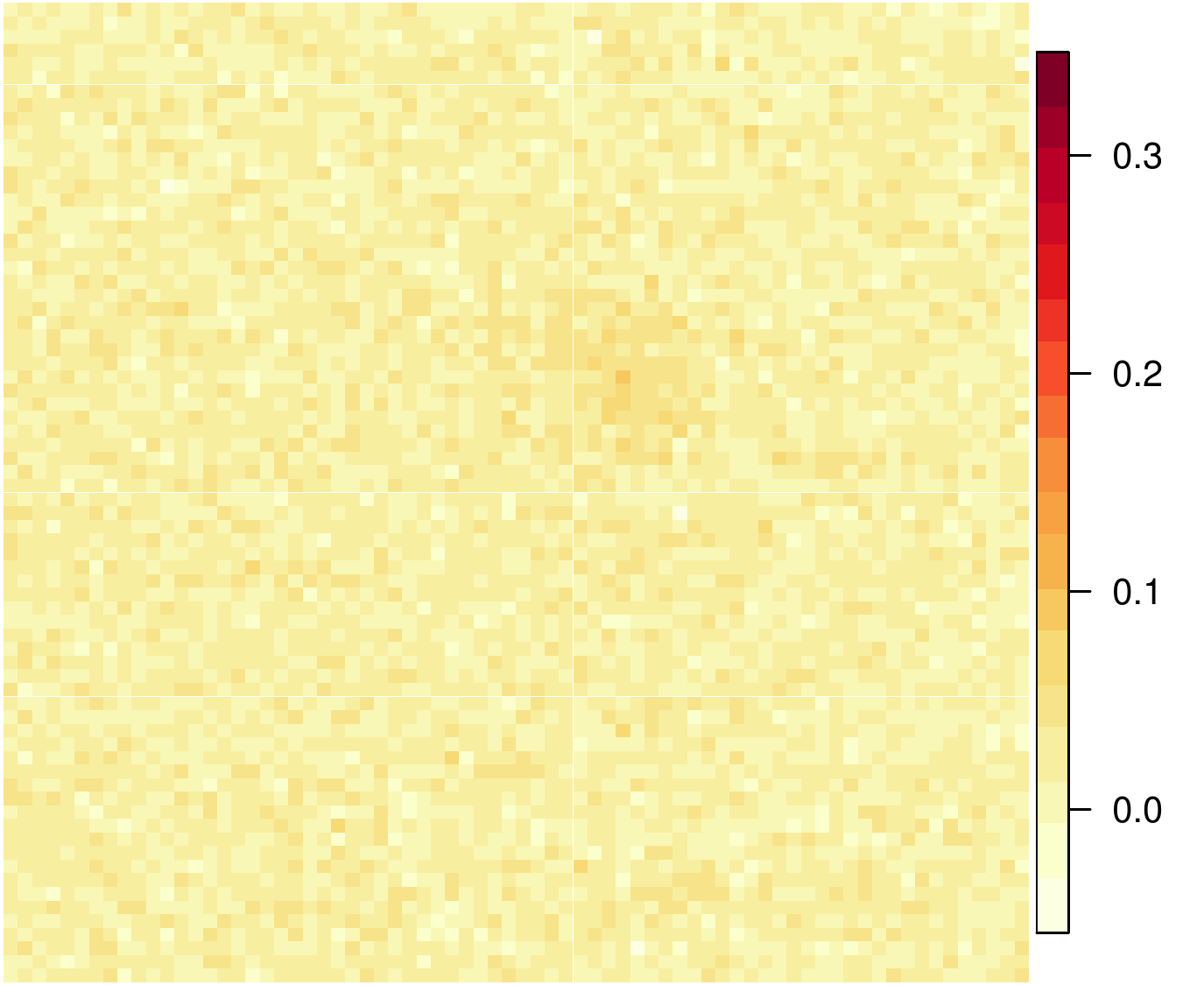}}
\subfigure[Noise image]{
 \includegraphics[width=0.31\textwidth]{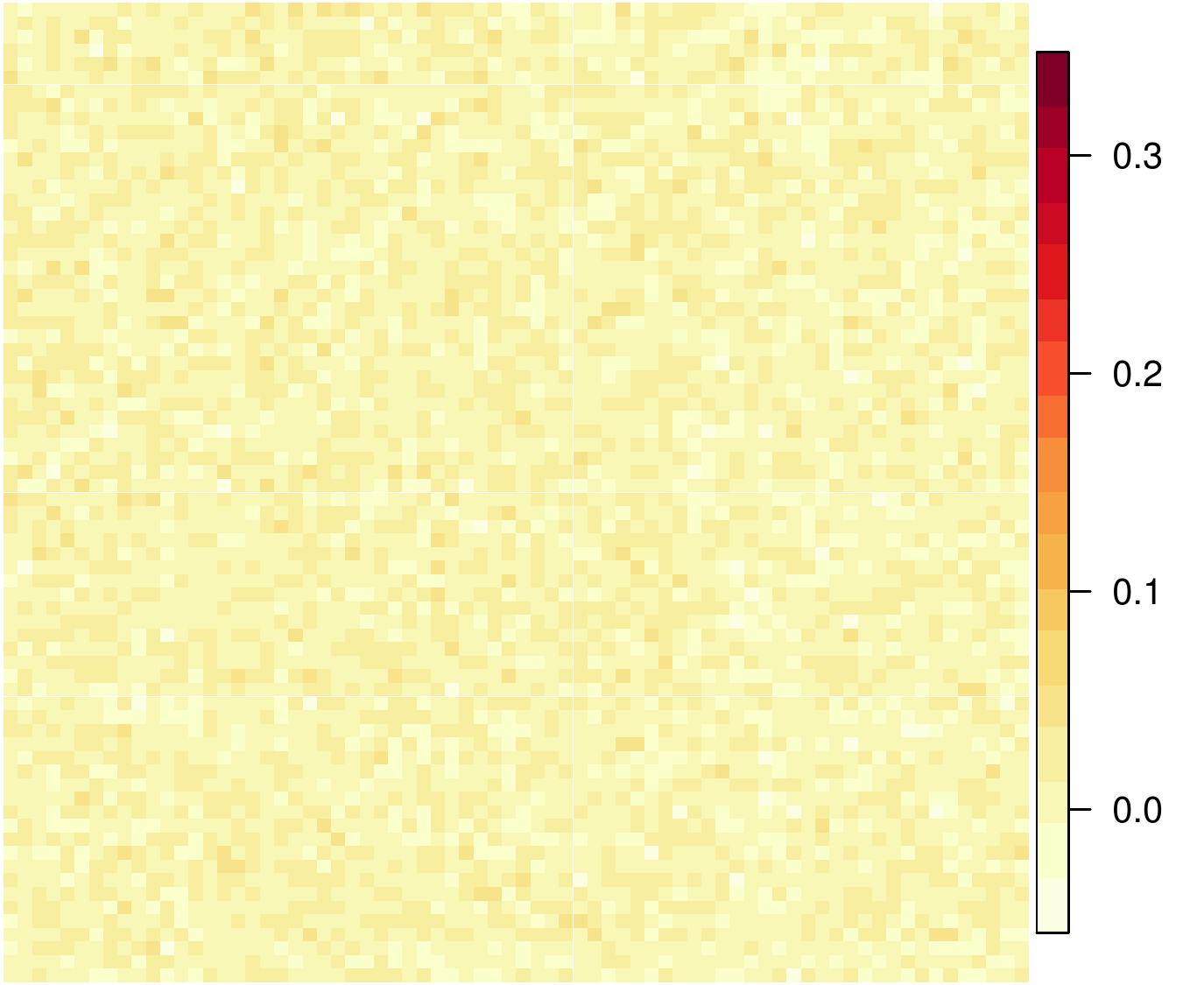}}}
\end{center}
\vspace{-0.25in}
\caption{Vibrothermography images on Ti-6Al-4V
  specimens with (a) a single large flaw, (b) a single small flaw and
  (c) no flaws.} 
\label{fig1}
\vspace{-0.15in}
\end{figure*}

Although the vibrothermography  technology is still in its infancy,
especially when 
compared to NDE images obtained using ultrasonic or radiographic
methods, a commonly-used pre-processing data reduction technique
(essentially eliminating the temporal dimension) is to use the image
frame with the largest   contrast (highest
signal)~\citep{lietal10,lietal11,gaoandmeeker12} before analysis. In
this paper, we consider these summary thermal images as the starting
point for our methodological development and analyses.
Figure~\ref{fig1} displays three sample thermal  images, each obtained
from a vibrothermographic time-course sequence of 150 image frames
collected on a titanium Ti-6Al-4V specimens with known flaw sizes (if
present). 
Figures~\ref{fig1}a and b display strong and weak signals in the
thermal  image as a result of a larger- and smaller-sized flaw while
Figure~\ref{fig1}c is a thermal  image of a specimen with no flaw and
thus is essentially an image only of the noise in image
acquisition. The three cases in Figure~\ref{fig1}  illustrate the 
challenges in determining the presence and size of a flaw and if
action is needed in terms of repair or replacement of the associated
part. In Figure~\ref{fig1}a, the flaw can be easily identified by a
trained operator. Distinguishing a case with a weak signal 
from one that only images noise is more challenging, however. Similar 
challenges also exist with respect to other NDE imaging
techniques such as those involving eddy currents~\citep{hasanzadehetal08},
ultrasound~\citep[see, {\em e.g.},][]{legendreetal01,ngetal13}
and radiography~\citep{wangandliao02}. 

\subsection{Background and Current Practices}
\label{background}
The development of analytical tools for vibrothermography image data
is still in an early phase, so we discuss current practices in the
context of other NDE imaging techniques, while noting that some of these
practices  have also carried over to vibrothermography. A common
approach, largely a cross-over from analysis methods used in
conventional ultrasonic NDE 
imaging, is to use the peak-amplitude method to detect a ``hot spot''
or region of elevated temperature readings (often after some initial
signal processing of the image). 
Here, an operator
visually finds the  hottest spot in an image and uses the pixel
intensity as the response for that image. Statistical models are then
fit to these responses and the result is used to claim detection of a
flaw when the observed response is higher than some threshold.   

A more refined approach~\citep{howardandgilmore94} to the above is
used in the analysis of  multi-zone ultrasound NDE images. Here, an
identified hotspot in a processed image is enclosed within two
visually-drawn rectangles in  
such a way that the pixels inside the inner rectangle have elevated
intensities while the region enclosed between the smaller and the larger
rectangles has  almost no high-intensity pixels. Thus, the 
inner rectangle conceptually encloses the signal region while the
outer (rectangular) frame represents the noise measurements. The pixels
in these two regions are used to calculate the SNR from which it is  
determined, as before, if the region enclosing the hotspot is higher
than the threshold, pointing to a possible flaw. 

There has been some attention paid to the issue of thresholding in NDE
imaging. \citet{howardetal98} proposed an SNR-based dynamic 
algorithm for detecting flaws in ultrasonic C-scan images. 
\citet{jansohnandschikert98}  described a statistical threshold
detection algorithm to interpret C-scan images of concrete 
elements by modeling the signal amplitude with a lognormal or,
alternatively, a Weibull distribution. Other work has focused on
improving the quality of image processing. For instance, 
\citet{chenandwang04} utilized independent components analysis (ICA)
for reducing speckle and to enhance edges of ultrasonic C-scan
images. \citet{lietal10} introduced  a three-dimensional~(3-D) matched
filter to enhance the SNR of the vibrothermography image sequence  and
also statistical methods for flaw detection using a matched filter
output. \citet{gaoandmeeker12} presented methodology for the
systematic analysis of image data from vibrothermography inspections,
based on principal components and  robust regression. 

An important issue in all methods, not addressed in the literature, is
the reliance on the human operator to  manually identify, after the
initial processing, the hotspot and, for \citet{howardandgilmore94}'s
approach, to also draw the two rectangles. While the human eye and the visual 
system have an unmatched ability in detecting and resolving many
situations, it is also true that operator fatigue and the low
probability of finding a flaw in most NDE inspections greatly impacts
detection accuracy, resulting in increased potential for missing an 
actual flaw. Thus, an approach which reduces this potential for
human-factor misses would be desirable. In this paper, we  develop a
automated statistical algorithm 
that  can identify images containing some evidence 
of the existence of a flaw. 
The objective behind this algorithm is to identify images
where the existence or otherwise of a flaw is easily and conclusively
established with a view to screening out images that require no 
further  evaluation by a trained expert. This will reduce the volume
of images required to be processed manually by a human inspector and
potentially decrease the probability of human-factor misses. 

As pointed out by the Editor, different aspects of the
issue of flaw detection in NDE images have similarities with the image
processing techniques of edge
detection~\citep{marrandhildreth80,rosenfeld84,bergholm87,
  gauchandpizer93,osullivanandqian94} and image
segmentation~\citep{rosenfeldandkak82-2,qiuandsun07,qiuandsun09}.
\citet{qiu05} points out that many of these techniques can be framed
in the context of jump regression. In edge detection, the objective is
to locate an object or object(s) in an image. Although many methods
are available, \citet{osullivanandqian94} provided an edge detection
algorithm using the {\em contrast statistic} to detect images of
single objects of arbitrary shape and size. For multiple objects, they
suggested iteratively locating boundaries one at a time. The contrast
statistic was applied to both emission computed tomography image data
as well as to simulations from a digitized phantom
experiment. \citet{osullivanandqian94} provided generalizations to
boundary detection in multi-channel and volumetric images, but the
performance of these methods in situations involving only noise (such
as would arise in most routine NDE inspections) is unclear. Image
segmentation forms a major sub-area within the ambit of image
processing, with the goal being to partition an image into 
different regions, each of which contain pixels that are similar to
others (with respect to some characteristic) in the same region, but
different in that same characteristic from pixels in another
region. Examples of such characteristics are the biochemical status
({\em e.g.}, in emission tomography) or composition ({\em e.g.}, in  Magnetic
  Resonance Imaging) of tissue or land types in images obtained  
by remote sensing. Our application in NDE imaging is, however,
centered solely on identifying possible hotspots in an image (in
actual NDE applications, the occurrence of a flaw is a rare event), 
and then determining  whether such a hotspot (if identified) is a
flaw or simply an artifact of noise. In this paper therefore, we develop an
automated statistical technique to identify potential hotspots in NDE
images.

 \subsection{Overview}

 The remainder of this paper is organized as
 follows. Section~\ref{methodology} develops our automated feature 
 detection algorithm and also incorporates extensions to account for 
 real-life special cases. We also develop  detection criterion and
 modeling strategies  relating the flaw size and NDE metrics extracted
 from our proposed automated approach. The performance of the
 proposed methodology is evaluated on vibrothermography and simulated
 datasets in Section~\ref{evaluations}. The paper concludes with some
 discussion in Section~\ref{discussion}. This paper also has an online
 supplement providing additional details on experimental
 illustrations, performance evaluations, and data analysis.  Sections
 and figures in the supplement referred to in this paper are labeled with
 the prefix ``S-''.

\section{Methodology}
\label{methodology}
\subsection{Preliminaries}
Let $\bZ \equiv \{Z(u,v):u=1,2,\ldots,n_1; v=1,2,\ldots,n_2\}$ denote
the observed $n_1\times n_2$ image with $Z(u,v)$ as the observed
intensity at the pixel with coordinates $(u,v)$. Let us denote the
true image signature using  
$\btau\equiv \{\tau(u,v):u=1,2,\ldots,n_1, v=1,2,\ldots,n_2\}$ and the
noise  with $\bepsilon\equiv \{\epsilon(u,v):u=1,2,\ldots,n_1,
v=1,2,\ldots,n_2\}$. Further, let $\mu$ be the systematic error at
each pixel coordinate. Then the model is given by 
\begin{equation}
\label{model}
  \bZ = \mu\bone + \btau + \bepsilon,
\end{equation}
where $\bone$ is the vector that takes the value 1 for all coordinates
and $\bepsilon$ is distributed 
according to some multivariate density $f(\cdot)$ with center zero in
each coordinate. 

We also review a few common NDE terms used in this paper. A
{\em crack}, {\em inclusion} or {\em porosity} is the name given to
different kinds of flaws that can affect the functioning or longevity of a
system component or part. The term {\em flaw} itself is generically
used to describe cracks, inclusions or porosities. Flaws are often
seeded into a test block or individual specimens that will be used in
experiments for evaluating performance of inspection systems; such
seeded flaws are called {\em targets}  because their existence and
location is known to the experimenters (but generally not to the
inspectors used in experiments). Finally, the term {\em indication} is
used to denote a hotspot, whether caused by a signal emanating from a true
flaw or a noise artifact.  

\subsubsection{Image Processing with Matched Filters}
\label{mfilter}
Image processing methods are used to improve SNR. One
approach, suitable for vibrothermography  
image data, is the matched filter~\citep{turin60,turin76} which we
review here.  Mathematically, given the model~\eqref{model}, the
matched filter output is, notionally, 
the two-dimensional~(2-D) convolution of the true signature $\btau$
with the observed $\bZ$ ({\em i.e.}, $\hat\btau = \btau\star\bZ$, with
$\star$ denoting the convolution operator). 
Note here that the true signature is not known (and is desired
to be estimated), so in practice, the matched filter is implemented by
convolving the known (or approximate) signature (also called a {\em
  template}) with the observed data  (consisting of signal plus noise)
to detect the presence of the signature in the output. 
A matched filter is an optimal linear filter in the sense that it maximizes 
the SNR in the presence of stationary white
noise when a signal's impulse response function (or signature) is
known. Even with spatially-correlated  noise, the matched filter
generally performs well. For more details, see, for example,
\citet{turin60}, \citet{turin76} or \citet[][Chapter 6]{engelberg07}.

 Matched filtering has its origins in communications and  signal processing
 and has been used as an image processing tool for various NDE
 imaging methods.  For example, when the approximate signal
signature is known, \citet{lietal10} used a 3-D matched
filter to process the sequence-of-image data 
from vibrothermography experiments for titanium Ti-6Al-4V specimens
containing fatigue  cracks. Also, in ultrasonic inspections, the high
background noise in titanium-alloy  parts makes detection
difficult, especially for small flaws. Matched  filter processing has
the ability to signficantly enhance SNR  and improve the probability
of  flaw detection. While there are several ways to construct a
matched filter, one approach that we have found promising utilizes the
Gaussian shape of the flaw signature. The basic idea -- modified from
\citet{lietal10} to apply to a 2-D setting -- places a
radially symmetric 2-D Gaussian kernel  at the
center of the imaging region. At the pixel $(u,v)$, this (discrete)
Gaussian signature is specified by $f(u,v)  \propto
\exp{[-2\sigma_\hbar^{-2}\{(u -     \bar u)^2 + (v - \bar v)^2\}]}$,
where the center of the imaging region is at $(\bar u, \bar v)$ and
the filter bandwidth is defined in 
terms of its full-width-at-half-maximum~(FWHM) of $\hbar$ pixels. (In
signal and image processing parlance, the FWHM of a filter is the range of the
interval formed by the two points where the signal attains half its
peak value. For a Gaussian filter with scale parameter $\sigma_\hbar$,
this translates to $\hbar = 2.355\sigma_\hbar$.) The choice of $\hbar$
is determined collectively by two criteria: (a) the Gaussian profile
is such that it vanishes on the boundary pixels and (b) the profile
provides reasonable resolution of the filtered output image in the
sense that is a moderate proportion of non-negligible-valued pixels in
the imaging grid.   

In the specific experiments reported in this paper, we have a imaging
grid of $30\times 30$-pixels.
The Gaussian shape used here is motivated by the thermal
Green's function~\citep{becketal92} which demonstrates that the
temperature profile of an impulse point-source is of Gaussian form. The
actual profiles in vibrothermography experiments vary from the
Gaussian form because the geometric shape of the heat source comes
from the locations of heating 
in the crack, and because the temperature profile is stepped rather
than an impulse~\citep{holland11}. Nevertheless, the diffusive nature
of the heat conduction 
\begin{figure*}[h]
\vspace{-0.15in}
\begin{center}
\mbox{
\subfigure[Strong signal]{
 \includegraphics[width=0.33\textwidth]{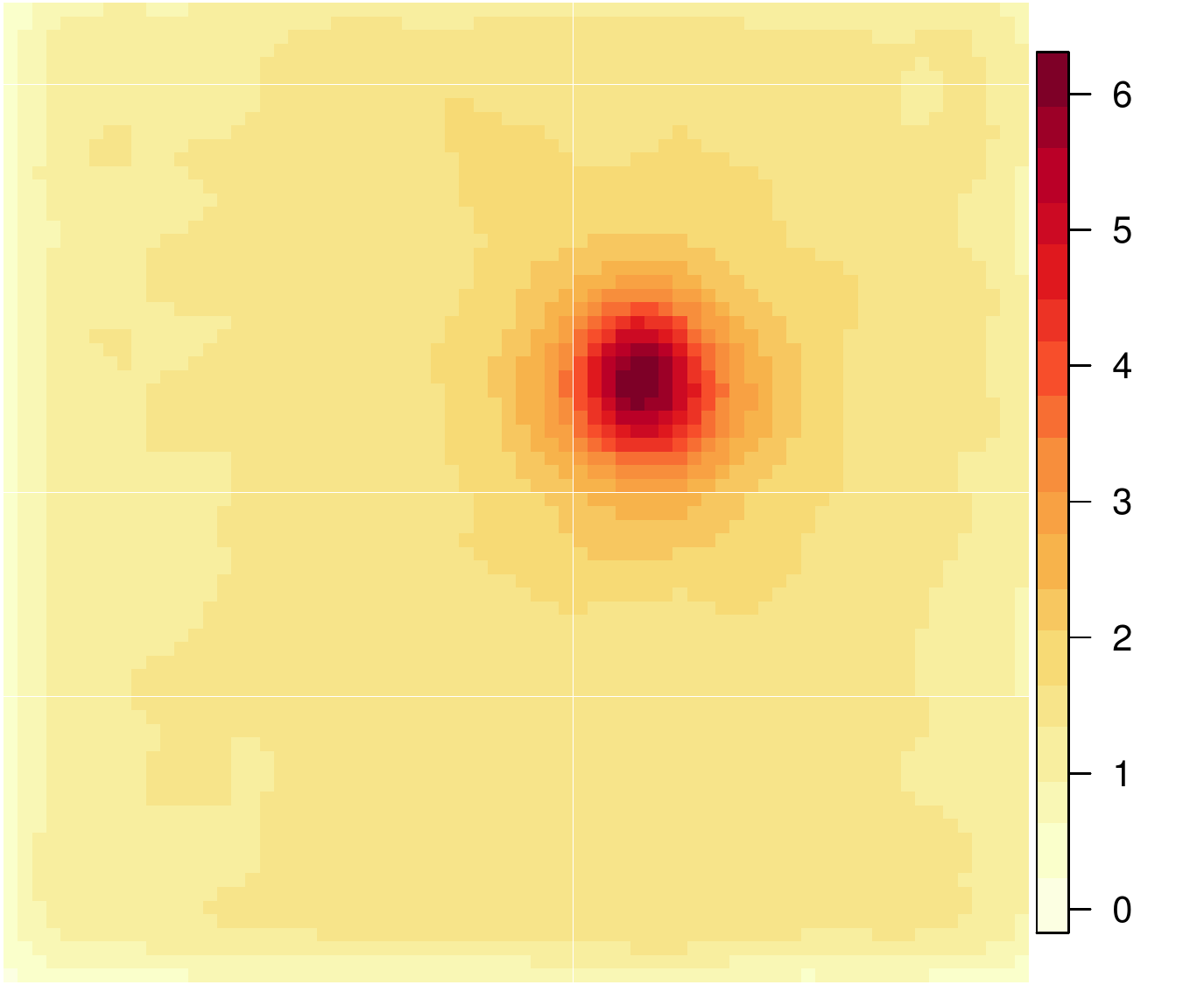}}
\subfigure[Weak signal]{
 \includegraphics[width=0.33\textwidth]{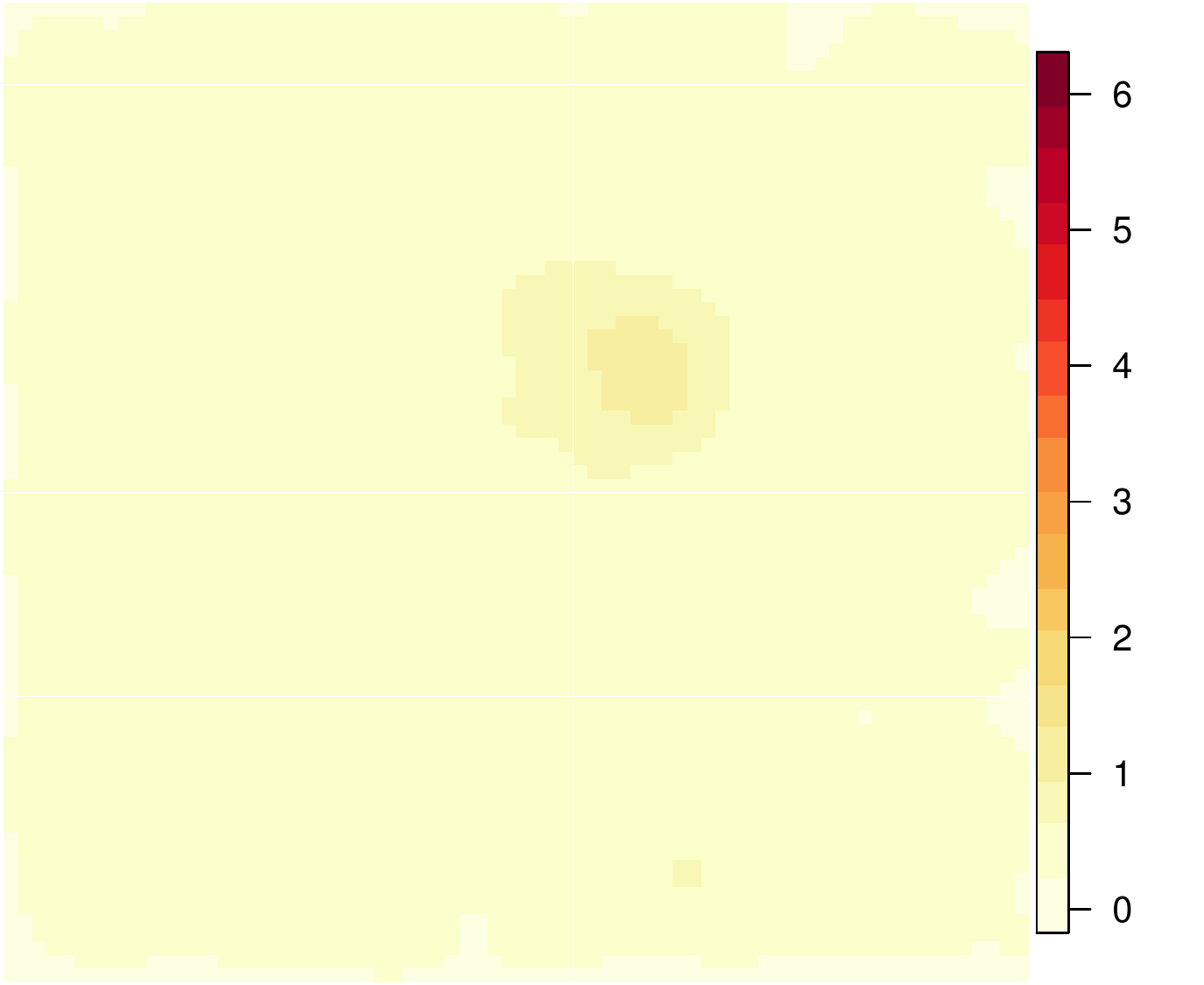}}
\subfigure[Noise image]{
 \includegraphics[width=0.33\textwidth]{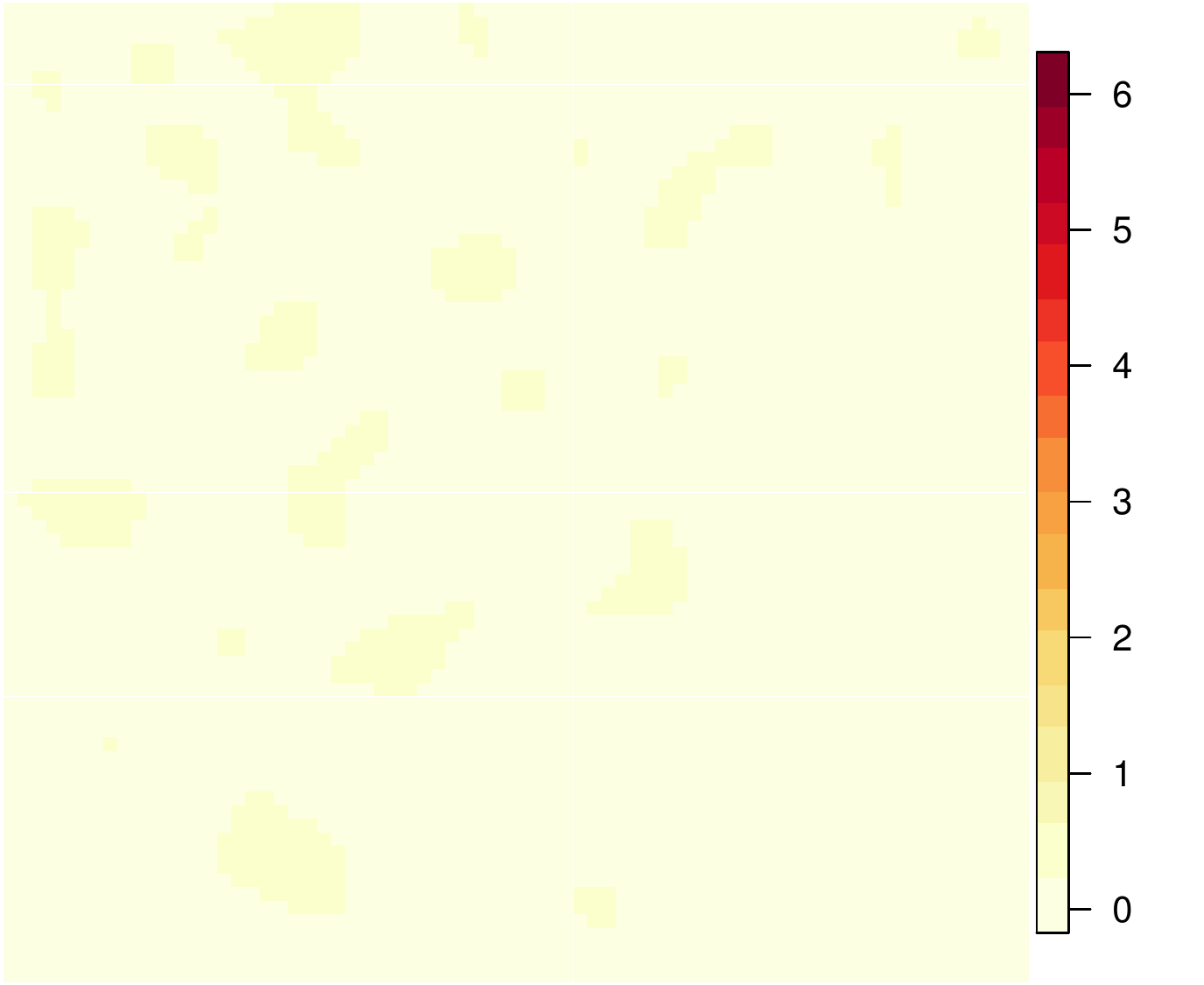}}}
\end{center}
\vspace{-0.3in}
\caption{Matched filter output for the vibrothermography image data
of Figure~\ref{fig1}}
\label{fig2}
\vspace{-0.2in}
\end{figure*}
equation means that the Gaussian shape provides a good approximation.
For our experiments, we have found  $\hbar = 4.71$ (equivalently
$\sigma_\hbar = 2$) to satisfy the criteria (a) and (b), though any
similar value could also have been used.  Figure~\ref{fig2} illustrates
the results of matched filtering  on the data of
Figure~\ref{fig1}.
There is substantial noise reduction (and SNR enhancement)
from the corresponding images of Figure~\ref{fig1}. However, it also
presents challenges, especially with regard to distinguishing material
with a small flaw (Figure~\ref{fig2}b) from a flawless
specimen~(Figure~\ref{fig2}c). Such processed images (of $\hat\btau$,
which for notational convenience, we henceforth denote as $\bY =
\{Y(u,v):u=1,2,\ldots,n_1;v=1,2,\ldots,n_2\}$)  
form the input for our proposed automatic statistical screening and
flaw detection algorithm. 

We conclude discussion here by noting that Wiener
filtering~\citep{jain89,engelberg07,gonzalezandwoods08} is often
also used in NDE imaging instead of matched filtering. In this paper
however, we focus development on matched-filtered images because this
filter has optimality properties with respect to SNR and because our
feature extraction and statistical modeling procedures  are  
SNR-based. However, our development is also generally applicable to  
Wiener-filtered images.  

\subsubsection{SNR Computations in NDE Applications}
\label{SNR}
The traditional definition of SNR in NDE applications compares the
ratio of the peak signal intensity to that 
of peak noise (after removing the estimated bias, {\em i.e.} the effect of $\mu$
in~\eqref{model}, for both terms). Though there are generally many
ways of defining SNR, we adopt this definition for consistency with
the application in this paper. Specifically, $\mbox{SNR} = (\check Y -
\bar e)/(\check e - \bar e)$, where 
$\check Y$ is the maximum (peak) value of the estimated signal (or 
hotspot value) of the processed image while $\check e$ and
$\bar e$ are the peak and average values of the estimated noise in the
pixels. We now adapt
\citet{howardandgilmore94}'s general strategy of identifying regions
having signal and noise (the inner rectangle and the outer frame for
their specific approach) to determine estimates for  $\check Y$,
$\check e$, or $\bar e$.

\subsection{Automated Feature Extraction Algorithm}
\label{algorithm}
This section develops methodology to automate the process of flaw
detection for NDE images, with special reference to vibrothermography image data. Our starting
point is the operator-controlled rectangle-drawing approach of
\citet{howardandgilmore94}. However, visual examination of large
collections of vibrothermography and ultrasonic  inspection images of
specimens containing flaws leads us to propose that the shape of most 
flaw signals can be described by ellipses. Thus, although one could adapt
nonparametric edge-detection methods \citep{osullivanandqian94,qiu05},
we choose to use a defined parametric shape in order to provide more
power to our detection algorithm, especially in low SNR
situations. Therefore, we assume that the signal region is 
elliptically-shaped. Generally a flaw signal can be covered by an
ellipse which we call the ``inner ellipse.''
We then draw another ``outer ellipse'' 
with identical center and orientation as its inner counterpart 
where the region between the two ellipses is expected to contain noise
pixels only. (These ellipses are elliptical cousins of
\citet{howardandgilmore94}'s inner and 
outer rectangles). We desire the inner ellipse to be as compact as
possible, so that it not only covers most of the signal pixels, but
also few or no noise pixels. After determining
the location and orientation of the inner ellipse, we use an
equal-area constraint ({\em i.e.}, the inner ellipse and the frame
have the same area) to determine the outer ellipse.

The use of the outer ellipse adjacent to the inner ellipse (which is
designed to encase the signal region) is because
the noise level can vary considerably within the inspected specimen.
Allowing the outer ellipse to have twice the area of its inner
counterpart ensures equal number of pixels in both the inner ellipse and
the adjoining frame, providing balance
in precision. However, it is certainly possible that a few pixels in
the frame have contamination from the signal, so we need to choose the
optimal elliptical regions to provide the best contrast (SNR) for
detection of a true flaw.

\subsubsection{Optimal elliptical regions}
\label{optimalelliptical}
\begin{figure*}[htbp]
\vspace{-0.05in}
\mbox{
\subfigure[Optimal ellipse]{
\includegraphics[width=0.24\textwidth]{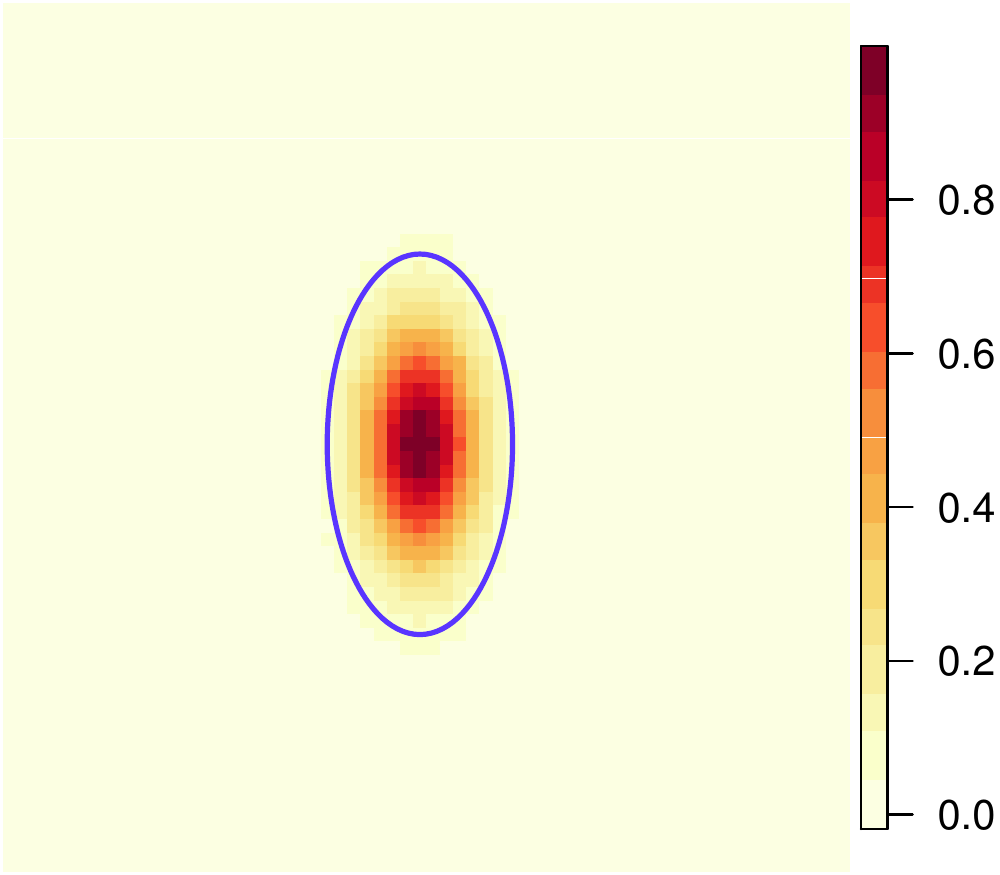}}
\subfigure[Over-sized ellipse]{
  \includegraphics[width=0.24\textwidth]{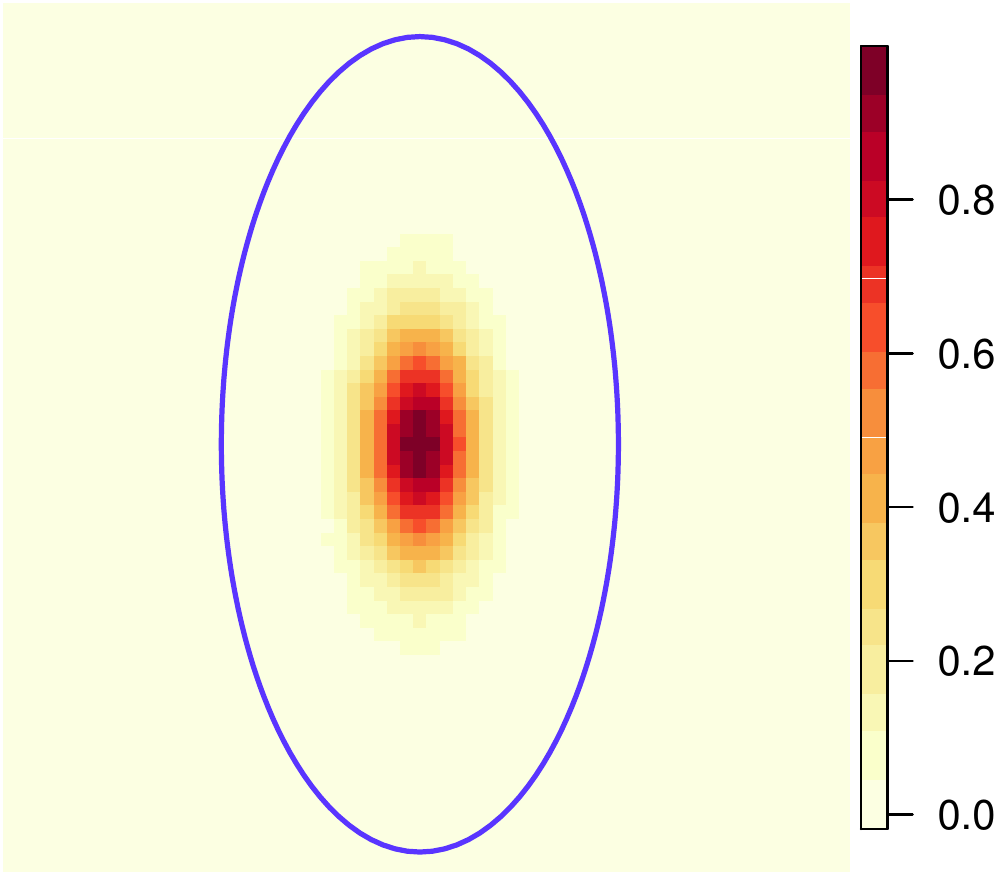}}
\subfigure[Under-sized ellipse]{
  \includegraphics[width=0.24\textwidth]{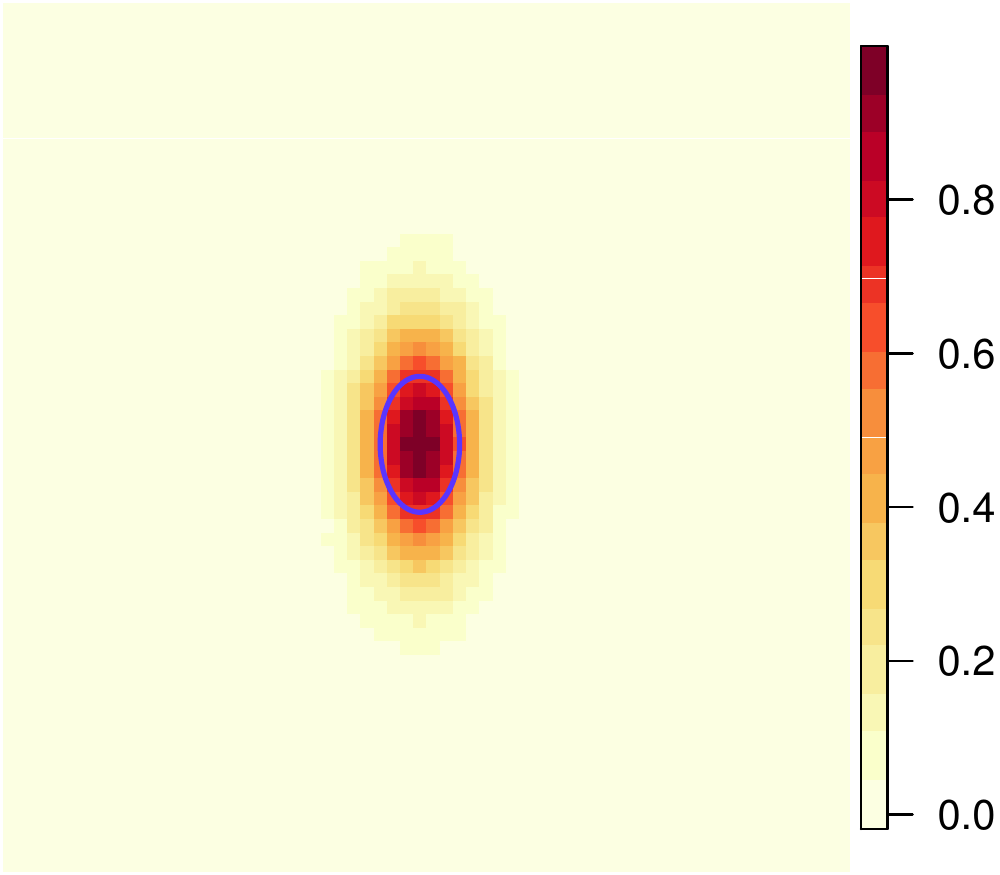}}
\subfigure[Misaligned ellipse]{
  \includegraphics[width=0.24\textwidth]{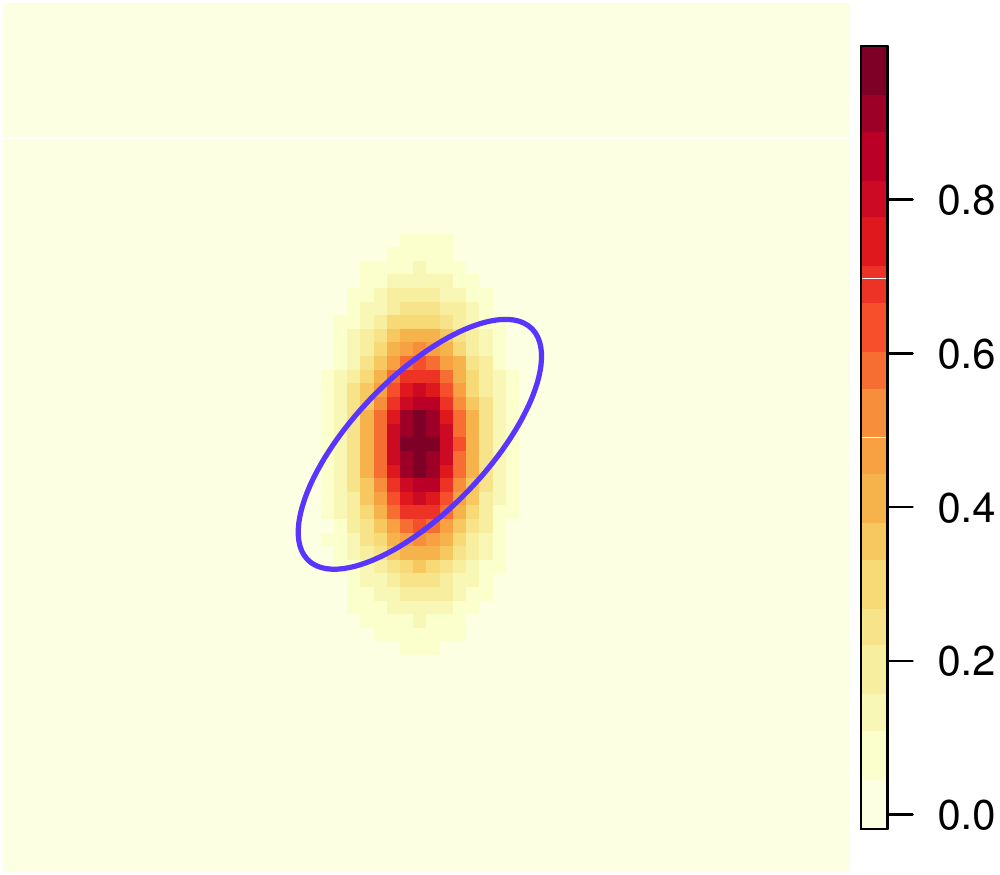}}}
\vspace{-0.1in}
\caption{(a) Sample elliptical signal region with drawn optimal ellipse matching
      the boundaries. (b) Over-sized, (c) under-sized and (d)
      misaligned ellipses that demonstrate the possibilities that
      justify the development of the specific $\mV(a,b,\theta)$ of
  Section~\ref{optimalelliptical}.} 
\label{allellipse}
\vspace{-0.15in}
\end{figure*}

 As mentioned above, we want the inner ellipse to
 be compact, containing as many signal pixels and as few 
 noise pixels as possible. Although human inspectors can manually draw
 such ellipses, we aim to provide an automated procedure
 that will increase throughput  and reduce  human factors
 variability.  We formulate the choice of the inner  
 ellipse in terms of an optimization problem, starting first with the
 case of only one indication in the  image. Then the main  steps of
 our algorithm are:
\begin{packed_enum}
\item \label{step1} {\em Determine the center of the signal:} Assume 
that the ellipse is centered at the ``hottest spot'' (pixel with
highest intensity $\check Y$) having coordinates $(u_\bullet
,v_\bullet)$. Denote the ellipse with parameters $(a,b,\theta)$ as
$\mE_i(a,b,\theta)$, where $a$ and $b$ are one-half of  
the  lengths of the major and minor axes and $\theta $ is
the angle between  the major and the horizontal axes.  

\item  \label{step2} {\em Determine the inner ellipse}:
  We introduce the concept of ellipse ``volume''
  $\mV(a,b.\theta)$ as a function of $(a,b,\theta)$ as follows:  For a
  given $\mE_i(a,b,\theta)$, we 
  define $\bY^i =\{Y(u,v):(u,v)\in \mE_i(a,b,\theta)\}$ and let
  $\bY^o=\bY\setminus\bY^i$ be the set of values $Y(u,v)$ outside the
  ellipse $\mE_i(a,b,\theta)$. Further, let $\hat\mu$ be an estimate of 
  $\mu$. Intuitively, we set $\hat\mu$ to be 
  equal to the mean of the intensities in $\bY^o$ ({\em i.e.}, 
  the mean pixel intensity outside $\mE_i(a,b,\theta)$). 
We subtract $\hat\mu$   
  from each element in $\bY$ to yield
  ``corrected'' intensities $\bY_c =\{Y_c(u,v)\}$, where
  $Y_c(u,v)=Y(u,v) - \hat\mu$. (Let $\bY^i_c$ and $\bY^o_c$ be the
  corresponding corrected intensities for pixels inside and outside
  the ellipse.) Intuitively, almost all of $\bY^i_c$ will have
  positive values, but the surrounding (contaminating) noise
  pixel intensities (in $\bY^i_c$) will be 
  scattered around zero (taking either positive or negative
  values). Then the volume is defined as
  \begin{equation} 
\label{voleqn} 
\begin{split}
\mV(a,b,\theta) & =\sum _{(u,v)\in\mE_i(a,b,\theta)} Y^i_c(u,v) \mI_{[Y^i_c(u,v) >0]} +\lambda 
\sum _{(u,v)\in\mE_i(a,b,\theta)} Y^i_c(u,v) \mI_{[ Y^i_c(u,v) \leq 0]} \\
&\equiv \sum_{(u,v)\in\mE_i(a,b,\theta)}Y^i_c(u,v) + (\lambda -1 )\sum_{(u,v)\in\mE_i(a,b,\theta)} Y^i_c(u,v) \mI_{[ Y^i_c(u,v) \leq 0]}
\end{split}
\end{equation} 
where $\mI_{[\cdot]}$ is the indicator function and 
$\lambda $ is a regularization parameter that can be used to
control the compactness of $\mE_i(a,b,\theta)$. 
For a fixed 
$\lambda $, the volume is a function of $(a, b, \theta)$. 
 Intuitively again, a large volume ({\em i.e.}, a large
 $\mV(a,b,\theta)$  value) corresponds to a compact inner ellipse
 containing mostly 
 signal pixels and only a small number of noise pixels. Therefore our
 goal reduces to maximizing the volume $\mV(a,b,\theta)$ as
 a function of $(a,b,\theta)$. 
The exact choice of $\lambda$ is left as a control parameter,
depending on the application, and the distinguishability of the true
flaw signal $\tau(u,v)$ relative to the standard error in the image.  

\item \label{step3} 
{\em Drawing the outer ellipse:}
Once the inner ellipse is drawn in Step~\ref{step2} , the outer
ellipse is drawn using the 
same center and orientation (as that of the inner ellipse) but
having twice the area (equivalently, the annular portion of the outer
ellipse has the same area as the inner ellipse). Within this
framework, the outer ellipse is specified to have axes lengths $a^* =
a+\Delta$ and $b^* = b +\Delta$. From the area restriction, we have
$a^*b^* = 2ab$ so that $\Delta =  [-(a+b) + \sqrt{a^2+b^2 + 6ab}]/2$. The 
outer ellipse thus drawn is used in obtaining the $\check e$ and $\bar
e$ for our SNR calculations.  
\end{packed_enum}
Our volume criterion has
similarities with the contrast statistic in edge detection of
 \citet{osullivanandqian94}. We now provide some theoretical basis
for the selection of $\lambda$ and  for the use of
$\mV(a,b,\theta)$. 
\paragraph{Theoretical justification of the volume criterion and
  guidance for selecting $\lambda$:}
 Suppose that the true signal region is  indeed  elliptically-shaped,
 with  all positive pixel intensities  inside the ellipse ({\em i.e.},
 $\tau(u,v) > 0$ for  $(u,v)\in\mE_i(a,b,\theta)$ and
 zero otherwise). Suppose also, for simplicity, that there is only one
 signal region in the image, and that $\bepsilon =
 (\epsilon_1,\epsilon_2,\ldots,\epsilon_n)$ is  modeled in terms
 of Gaussian  white noise, that is,  $\bepsilon\sim N(\bzero,
 \varsigma^2\bI)$, where $\bzero$ is the vector having zeroes in all
 coordinates and $\bI$ is the identity matrix of appropriate order.
  Given $\mE_i(a,b,\theta)$, one may use $\hat\mu =\bar Y^o =
\bone'\bY^o/n_o$ to estimate $\mu$. In our implementation, we
compared results using the sample mean $\bar Y^o$ as well as the
sample median $\tilde Y^o$ with virtually indistinguishable
performance so we only report  results using $\bar Y^o$. 
In any case, both $\bar Y^o$ and $\tilde Y^o$ are consistent
estimators for $\mu$ and since $n_o$, the number of pixels outside
$\mE_i(a,b,\theta)$ is anticipated to be large, large-sample
arguments hold. Then, $\bY_c^i$ and $\bY_c^o$ are the corrected
intensities inside and outside the ellipse after accounting for the
systematic estimated bias. In particular, in terms of $n_o$, each
$Y_c^i(u,v)$ is 
normally  distributed with mean $\tau(u,v)$ and 
variance $\sigma^2$. 
From~\eqref{voleqn} 
we have 
\begin{equation}
\hspace{-1mm}  \E[\mV(a,b,\theta)] = \sum_{(u,v)\in\mE_i(a,b,\theta)} \left\{\tau(u,v)
    + (\lambda-1) \left[\tau(u,v) \Phi\left(-\frac{\tau(u,v)}{\sigma}\right) -
      \sigma\phi\left(-\frac{\tau(u,v)}{\sigma}\right)\right]\right\}
\label{expvol}         
\end{equation}
where $\Phi(\cdot)$ and $\phi(\cdot)$ are, respectively,
the cumulative distribution function and the probability density
function of the standard normal random variable $Z$. The multiplier for
the term $(\lambda-1)$ inside the summation of \eqref{expvol} follows
from \eqref{voleqn} since 
$  \E\{Y_i^c(u,v)\mI_{[Y^c_i(u,v)<0]}\} =
  \sigma\E[Z\mI_{[Z<-\tau(u,v)/\sigma]}] +\tau(u,v)\Phi(-\tau(u,v)/\sigma) $
  and then noting that for any $\zeta\in\R$,
$    \E\{Z\mI_{[Z < \zeta]}\} = \frac1{\sqrt{2\pi}}\int_{-\infty}^\zeta
  z\phi(z)dz =   \phi(\zeta),$
upon substituting $w=z\exp{(-{z^2}/2)}$ in the integrand. For ease of
presentation, let us use $C_\sigma(t) = t + (\lambda-1) \left[t 
  \Phi\left(-{t}/{\sigma}\right) -
  \sigma\phi\left(-{t}/{\sigma}\right)\right].$ 
Thus, $C_\sigma(\tau(u,v))$ is the contribution at the $(u,v)$th pixel to
the summation in \eqref{expvol}. Note that $C_\sigma(t)$
is a 
monotone increasing function in $t$ with first derivative
$C_\sigma'(t) = 1+(\lambda-1)\Phi(-t/\sigma)$. For any pixel outside
the true signal  
region, $\tau(u,v) = 0$, so that its contribution
$C_\sigma(\tau(u,v))$ to the summation in~\eqref{expvol} is
$-(\lambda-1)\sigma/\sqrt{2\pi}$ which is negative as long as $\lambda >
1$. For pixels with positive $\tau(u,v)$, the sign of
$C_\sigma(\tau(u,v))$ depends on the exact value that $\tau(u,v)$
takes relative to $\sigma$ and $\lambda$. Let $\xi =
\min\{\tau(u,v)/\sigma:\tau(u,v) > 0\}$ and let
$\lambda_\xi$ be the
root of the equation
\begin{equation}
h_\xi(\lambda) = \xi\left[1+(\lambda-1)\Phi(-\xi)\right] -
(\lambda-1)\phi(-\xi),
\label{lambdaeqn}
\end{equation}
which is obtained by dividing $C_\sigma(t)$ by $\sigma$ and setting
$\xi=t/\sigma$. 
Then, $\forall\lambda \in (1, \lambda_\xi)$, 
$C_\sigma(\tau(u,v)) > 0 $  for all $(u,v)$ in the signal
region. We illustrate our ideal ellipse using a
simulated image (after processing with
a matched filter) in Figure~\ref{allellipse}a. Note that
after allowing for the effect of pixelization, the optimal
ellipse is perfectly aligned with the boundary of the signal
region. Further, all terms in the
summation of~\eqref{expvol} are positive because all pixels with a
negative $C_\sigma(\tau(u,v))$ are outside the drawn ellipse and so
are excluded from the calculations. Also, the assumption
of no more than one signal region means there are no 
pixels outside the ellipse with 
$C_\sigma(\tau(u,v))>0$. We now show that, under these assumptions,
there are three overarching cases that we need   to consider (because all
other cases are 
subsumed within these three possibilities), but all these cases result
in a lower $\mV(a,b,\theta)$ on the average. 
\subparagraph{Case 1 - Over-sized ellipse drawn beyond the boundary
  of the signal region:} Here, as displayed in Figure~\ref{allellipse}b, the 
drawn ellipse covers all of the signal region, but also contains  a
substantial amount of area with no signal. In this case, the positive
terms that contribute to the summation in the optimal ellipse of
Figure~\ref{allellipse}a get depleted by the (noise) pixels outside the signal
region that are included inside the ellipse because the contribution
$C_\sigma(\tau(u,v))$ of  these noise pixels to $\mV(a,b,\theta)$ is
negative. Thus, in expectation,  
the $\mV(a,b,\theta)$ is less than that for Figure~\ref{allellipse}a. 
\subparagraph{Case 2 - Under-sized ellipse drawn inside the signal
  region:} Figure~\ref{allellipse}c illustrates the case where 
the drawn ellipse is inside the signal region but some signal
pixels lie outside the ellipse and are considered as noise. Here, two
factors reduce $\E[\mV(a,b,\theta]$. Primarily, there 
are fewer positive terms $C_\sigma(\tau(u,v))$ 
included in the summation than for the case with
Figure~\ref{allellipse}a because pixels in the 
signal region but outside the drawn ellipse have been excluded from
the summation. Further, the estimate $\bar Y^i$ of $\mu$ is biased
upwards because it includes pixels with $\tau(u,v) >
0$. Consequently, $\E[Y^i(u,v)] \leq \tau(u,v)$ and because 
$C_\sigma(t)$ is a monotonically increasing function, 
each of the included contributions to the summation in~\eqref{expvol} is
also potentially reduced. 
Thus, we also get a lower $\E[\mV(a,b,\theta)]$ than in
Figure~\ref{allellipse}a. 
\subparagraph{Case 3 - Ellipse drawn with misaligned axis:} The drawn
ellipse is misaligned, which means that the orientation of its principal
axes is at variance with that of the (elliptically-shaped) signal. As
a result, some signal pixels lie outside the ellipse, while some
others  that are not part of the signal region are included inside
it. Thus, the terms in the summation of~\eqref{expvol} include some 
$C_\sigma(\tau(u,v))$-values that are negative (these are the noise pixels
with $\tau(u,v) = 0$) while also excluding some pixels for which
$C_\sigma(\tau(u,v))$ has a positive contribution (signal region
pixels that are outside the ellipse, resulting in the same reduction of
$\E[\mV(a,b,\theta)]$ as in Case~2 above). In this case also,
$\mV(a,b,\theta)$ is expected to be lower  than in the case where the
ellipse matches the boundary of the signal region
(Figure~\ref{allellipse}d).    

The discussion here provides some theoretical grounding for the use of
$\mV(a, b, \theta)$ as an objective function to be maximized in terms
of the inner ellipse parameters $(a,b,\theta)$. We have shown that for
over-sized, under-sized or disoriented
ellipses, the volume criterion is,  on the average, smaller than that
for the case where the ellipse tracks the boundary of the signal
region. Thus, maximizing $\mV(a, b, \theta)$ as a function of  
$(a,b,\theta)$ is a necessary step to  obtain the optimal
ellipse. Finally, we restrict our search for $(a,b,\theta)$ to be
in a closed set to guarantee the existence of a maximum for a given value of
$\lambda$.

A reviewer has asked about the algorithm used to obtain the
global maximum. Like with many other iterative optimization algorithms,
convergence to a maximum depends on the initializing values. We take
some care in choosing these initializers by first using a
3-D grid of starting  values for $(a, b, \theta)$ in
$(10\!\times\!10\!\times\!10)$-grid and then, evaluating $\mV(a,b,\theta)$
at each grid value. The combinations 
providing the five largest value of \eqref{voleqn} are each run for
ten ``short'' iterations. The best performer at the conclusion of
these ten ``short'' iterations is then used to start the optimization
algorithm and run to convergence. 

Our initialization algorithm and convergence assessmment is a hybrid
adaptation of the {\em em-EM}~\citep{biernackietal03} and  and {\em
  Rnd-EM}~\citep{maitra09} algorithms used in the context of
intializing the Expectation-Maximization (EM)
algorithm~\citep{dempsteretal77,mclachlanandkrishnan08} for parameter
estimation in Gaussian mixture models. In {\em em-EM}, the EM
algorithm  is initialized at several (random) initial values and then
run to lax convergence with each initializer for a fixed total number
of iterations. The solution providing the highest loglikelihood value
at lax convergence is then run to strict convergence. The {\em Rnd-EM}
algorithm trades off the lax convergence steps in {\em em-EM} by
eliminating it in favor of a (large) total number of initializers at
which the loglikelihood is evaluated, with the solution that produced
the highest loglikelihood value then run to strict convergence. Both
these methods have been shown to be competitive
initializers~\citep{maitraandmelnykov10} with no clear winner so we 
adapt a hybrid scheme of both in initializing our optimization
algorithm. As a check, we have reviewed contour plots of the volume
function for several sample vibrothermography datasets and found no cause for
concern about the existence and identification of the global maximum.

\subsubsection{Effect of $\lambda$}
The objective function~\eqref{voleqn} depends on the regularization
parameter $\lambda$: Figure~\ref{lambdaeffect} illustrates that a
larger $\lambda$ value leads to smaller ellipses and vice-versa. We now
invoke~\eqref{lambdaeqn} in developing guidelines for choosing 
$\lambda$. 
The root of~\eqref{lambdaeqn} is given by
$
\lambda_\xi = \xi/[\phi(-\xi) - \xi\Phi(-\xi)]+1
$
so that the choice of $\lambda_\xi$ depends on $\xi$
which is the minimum
intensity  of the signal region relative to the noise standard
deviation $\sigma$. Consequently, $\xi$ can be viewed as the minimum
value that a flaw signature can be expected to have in order to ensure
a high POD. This is context-dependent and should ideally
be  set according to a standard determined by the 
particular application. Our discussion above has shown that  we have
$C_\sigma(\tau(u,v)) > 0$ as long as
$\lambda\leq\lambda_\xi$. However, given that only flaws of intensity
greater than $\xi\sigma$ are expected to be of importance in our application, we
recommend setting $\lambda\equiv\lambda_\xi$. We emphasize that our
recommendation of $\lambda_\xi$ is only a guideline: our derivation
above has invoked 
the distributional assumption of Gaussian white noise, which may not
be accurate  for the processed image. We have found, however, that
the exact value of $\lambda$ around a given $\lambda_\xi$ does not
appreciably alter the results. In particular, there is not much
difference in results for $\lambda_\xi$-values obtained over a range
in $(\lambda_{z_{0.95}},  \lambda_{z_{0.975}})\equiv
(79.73, 208.49)$. For our experiments 
reported in this paper, we  have used $\lambda = 100$, simply as a
matter of choice. 
\subsubsection{Some extensions of our algorithm}
\label{extensions}
\begin{figure}[h]
  \vspace{-0.1in}
\begin{center}
  \mbox{ 
    \subfigure[]{\includegraphics[width=0.49\textwidth]{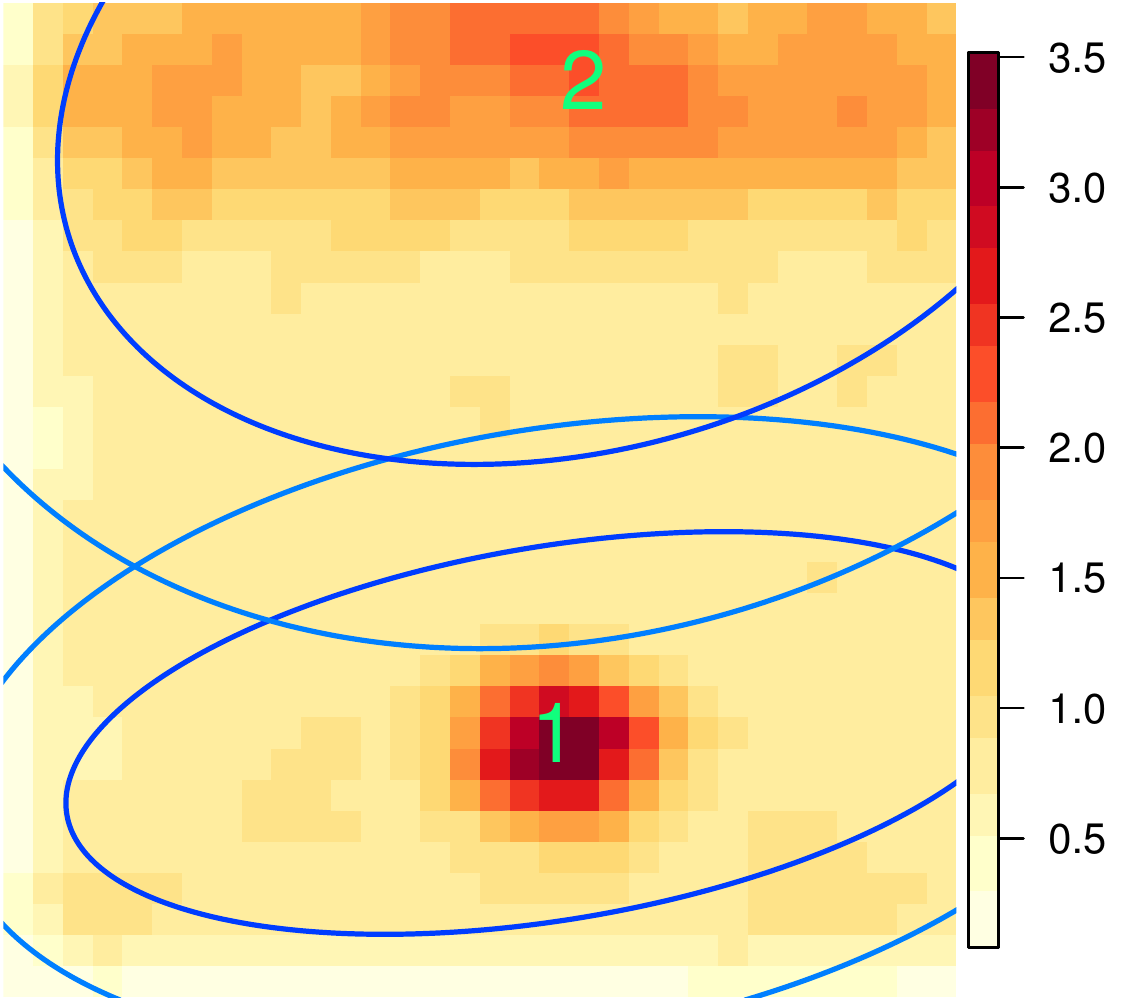}}}
  \mbox{
  \subfigure[]{\includegraphics[width=0.49\textwidth]{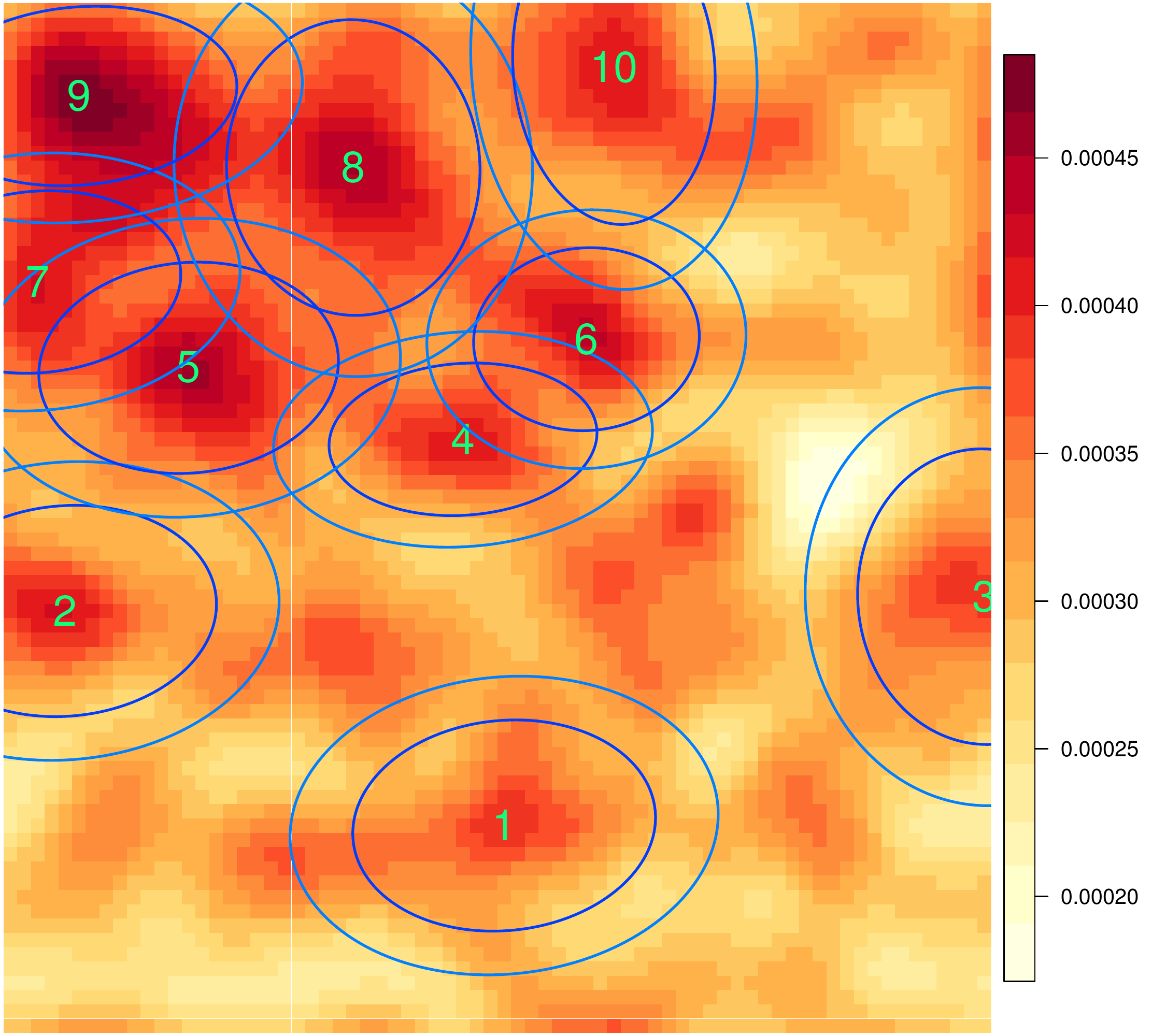}}}
  \vspace{-0.15in}
  \caption{Detection of multiple potential targets in (a) an image
    ultrasound and (b) a vibrothermography image of a noise specimen
    using our proposed algorithm.}
  \label{2targets}
\end{center}
\vspace{-0.25in}
\end{figure}

Our algorithm has so far been developed for one (elliptical)
signal region. Some NDE imaging applications could have
multiple hotspots while some others could have a pair
of hotspots ({\em e.g.}, from a crack with two tips) generated from a
single flaw. We now extend our algorithm for these  cases.
\paragraph{Extension allowing for multiple flaws:}
Figure~\ref{2targets}a displays an ultrasound image after processing
with a matched filter to illustrate a case with potential
non-negligible probability of having multiple flaws.
Briefly, ultrasound images are used to find near-surface and
subsurface flaws in materials by transmitting ultrasonic waves through
them. Specifically, Figure~\ref{2targets}a is an image of a 
synthetic inclusion forging disk (known as SID) as described 
in Section 6.2 of~\citet{margetanetal07}. This particular  SID
contains numerous types of synthetic  hard alpha (SHA) inclusions and
flat bottom holes (FBHs) of different known sizes. Additionally, for
titanium and other similar noisy materials with large grain
boundaries, there is also noise as a consequence of banding, leading
to a combination of low- and high-noise regions.  In the illustration,
the disk is known to have an  SHA as well as signal from a high-noise
region of the SID.
Our objective  is to extend our algorithm to make it
possible to identify a flaw in the presence of multiple hotspots 
(the  SHA and the banding noise in this example) and to evaluate their
SNRs for analysis. We do so by incorporating the additional steps:
\begin{packed_enum}
\item First, we detect centers for all possible 
  indications under the assumption that each  signal region is elliptical, with
  intensity decreasing away from the center of the region where it is
  highest. For each such region, draw a square of $3\times 3$ 
  pixels and define it as  a  possible flaw if the intensity at its 
  center is not smaller than at all its surrounding 8 pixels. By
  looking at  all such $3\times 3$-pixels moving windows in  the image, we
  can identify all possible indications.  
\subparagraph{Comment:} The use of the $3\times 3$-pixels window in the
manner described here is designed to filter out potential candidates
that are simply noise pixels with high observed intensities. If the
center of a $3\times 3$-pixels window has the highest intensity in
relation to all the other eight pixels that surround it, then it is
much more likely to be the center of a true flaw. 
\item 
For each indication detected in the previous step, quantify the peak
amplitude ({\em i.e.}, the pixel intensity at the center). Calculate
the  scaled amplitude as the ratio of the amplitude to the maximum
amplitudes for each such candidate indication.
\item
  \label{multi.ind}
Each of the indications obtained in the previous steps are candidate
flaws whose exact status needs to be determined. A  number of these
candidates are noise artifacts with virtually no chance of being a
true flaw. We reduce computation time 
(in subsequent evaluation and processing stages) by eliminating such
candidates. Specifically, 
we eliminate all those candidates  with scaled amplitude less than a
pre-set threshold $\varrho\in(0, 1)$. In the experiments of this
paper, we set $\varrho=0.9$.  
\item 
  As a final step, maximize the volume~\eqref{voleqn} on each
  indication selected in Step~\ref{multi.ind} to draw the optimal
  inner   and outer ellipse for feature extraction on the selected
  indications. Thus, we get a set of indications, each of which is a
  potential flaw that needs further analysis for accurate determination.
\end{packed_enum}
We refer back to Figures~\ref{2targets}a (and b) to illustrate the performance
of our algorithm in detecting multiple flaws. Our algorithm identified
two hotspots indicated by ``1'' and ``2'' in Figure~\ref{2targets}a in
the decreasing order of their scaled 
amplitudes.  (The inner and outer ellipses for SNR calculations are
also drawn around each indication.) Indication  ``1'' in the image 
corresponds to a flaw  arising out of an SHA while indication ``2''
corresponds to the banding background  titanium  noise mentioned
earlier. Figure~\ref{2targets}b illustrates performance of our
algorithm on a noise specimen in vibrothermography. Note that the fact
that there is no true signal in an image from a noise specimen means
that there is greater potential 
for identification of multiple indications, and indeed
Figure~\ref{2targets}b does identify multiple possible targets. (Once
again, we have drawn the inner and outer ellipses around each
potential indication.) We conclude our 
discussion of this illustration by noting that although two candidate
flaws were selected in Figure~\ref{2targets}a using the steps above,
only indication ``1'' gives a relatively high SNR of 7.90. The SNR for
indication ``2'' is  2.03, suggesting that it is likely a noise
artifact. For Figure~\ref{2targets}b, we identified 10 possible
targets but these all had SNRs below 2, indicating that it is likely
that all these are noise artifacts. We provide a more 
formal approach to deciding  on the detection limit for SNR-based
metrics in Section~\ref{detection}, noting here only  that the extension of
our algorithm can successfully  accommodate multiple flaw detection.
 
\begin{figure*}[h]
\mbox{
  \subfigure{\includegraphics[width=0.33\textwidth]{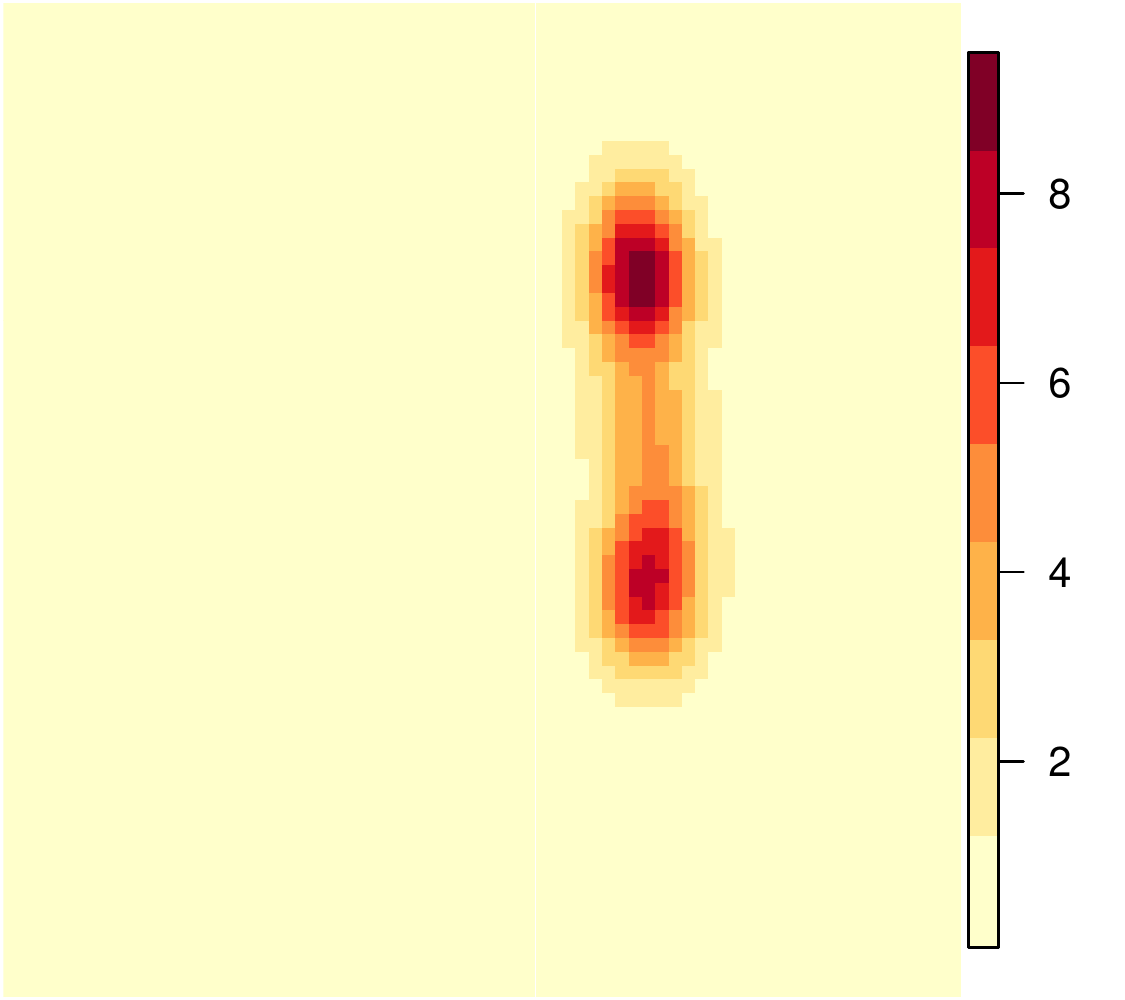}}
    \subfigure{\includegraphics[width=0.33\textwidth]{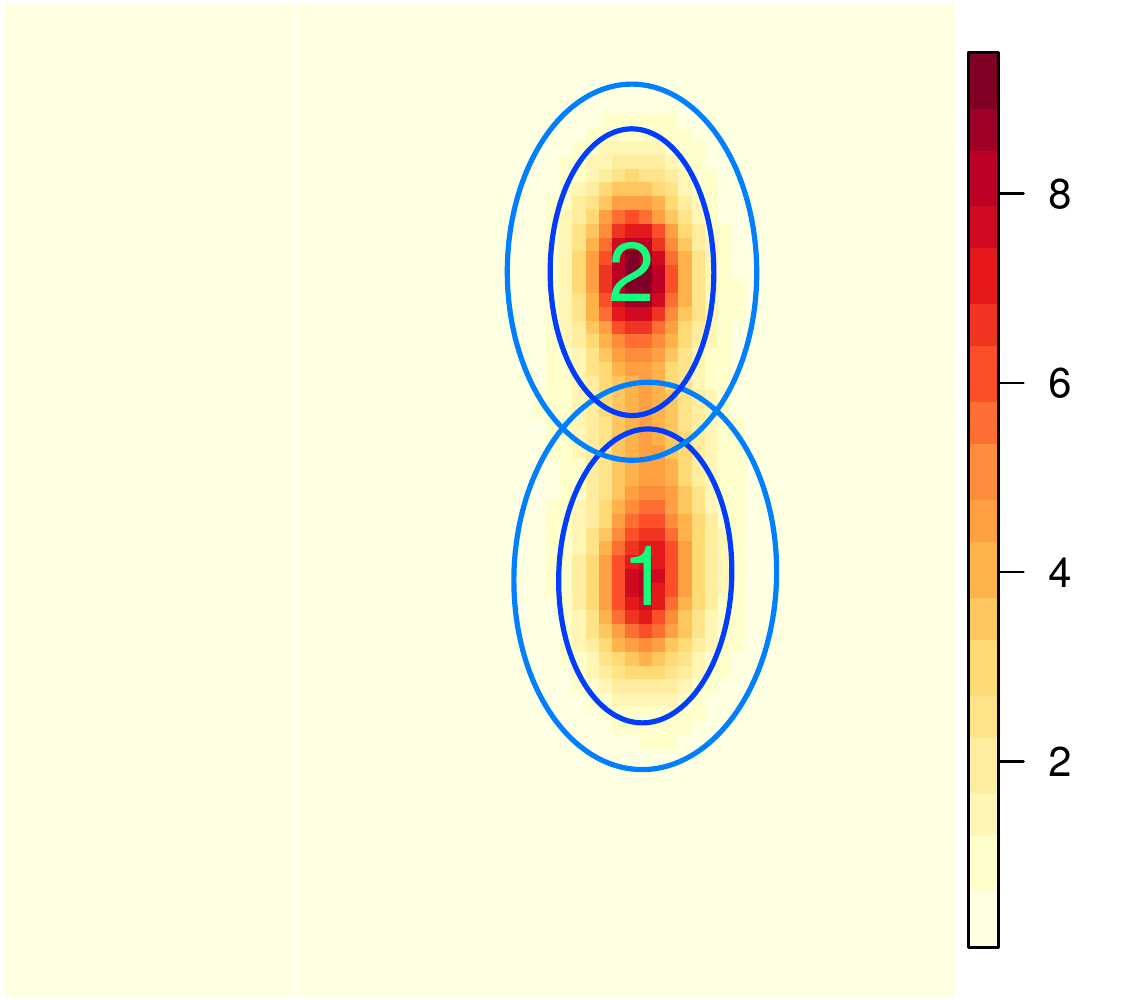}}
    \subfigure{\includegraphics[width=0.33\textwidth]{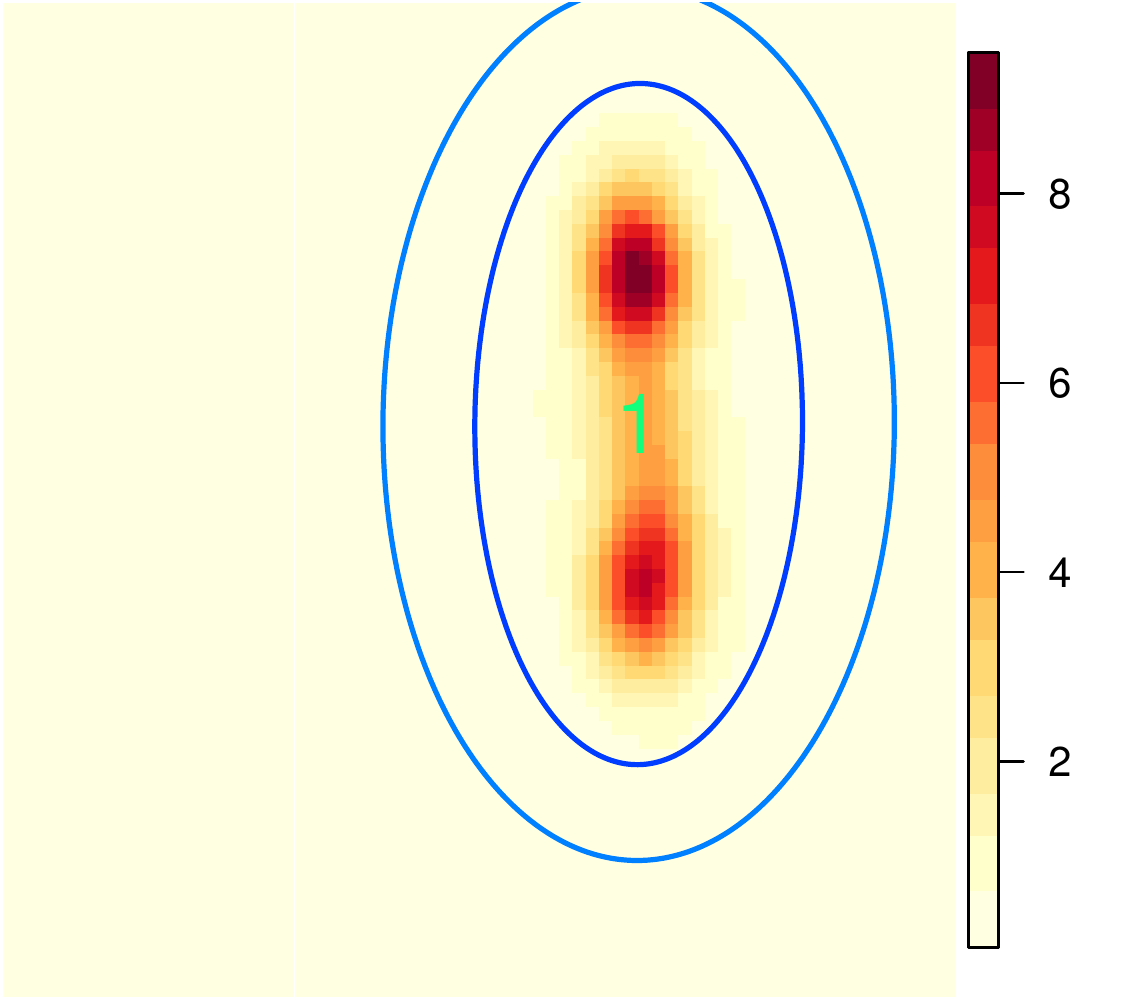}}}
\caption{(a) Processed vibrothermography image with two hot tips in
  one indication region and illustration of (b) algorithm for   detecting
  multiple hotspots and (c) its extension for deducing that  
  the two hotspots are in reality from one signal souce.}
\label{pairhotspots}
\vspace{-0.1in}
\end{figure*}

\paragraph{Extension to allow for signal in the form of a pair of hot
  regions:}
In some applications, including vibrothermography, flaws may develop
and be imaged in the form of two indications that are close to
each other with a bright region in between, as shown in the
vibrothermography image in Figure~\ref{pairhotspots}a of a 
seeded flaw after processing with a matched filter.
In this particular example, there is only one crack but heat is
generated from the extremal crack tips.
For such cases, although we may be  able to draw a set of
ellipses for each of the two hotspots, the SNR's may be low. This is
because the outer elliptical frames contain some high-intensity
pixels, thus enhancing $\bar e$ and attenuating the SNR. In order to
address this situation, we can consider two peaks as belonging  to a
single indication and create a new center as the midpoint of centers
of the two original ellipses. Then we can draw a new set of ellipses covering
both signal indications and extract useful metrics. The additional 
algorithmic steps are as follows:
\begin{packed_enum}
\item
 First, for any pair of hotspot indications detected using the
 methodology and extension described above, check if the two
 centers are close to each other.  The threshold to detect
 closeness can be pre-specified in terms of 
 the  number of pixels, corresponding to  the largest expected
 flaw  size. 
\item
 If the two centers are  close, with corresponding SNR's that are
 smaller than some pre-specified SNR threshold ({\em e.g.}, one 
 may adopt 2.5 as the SNR threshold following the  standard used for
 ultrasonic flaw detection in titanium parts), use the
 midpoint of the two   centers as the new center, and draw another
 (larger) set of ellipses that covers  both indications ``1'' and ``2''
 (both hotspots are now considered to be part of the same
 indication). If the new SNR (corresponding to the larger ellipse) is
 greater than both the original SNRs, we use  the new region (larger
 ellipse) to replace the original two indication regions (smaller
 ellipses). 
\end{packed_enum}
Figures~\ref{pairhotspots}b and c illustrate the application of the original  
algorithm and the above  extension on the image of
Figure~\ref{pairhotspots}a.   Using the original algorithm without 
pairwise correction, two sets of ellipses can be nicely drawn, but
these have relatively low SNR values (2.44 for indication ``2''  and 2.12
for indication ``1''). These SNR's are below the standard
threshold SNR  of 2.5 so we cannot claim these indications
individually as flaws, even though the existence of the flaw is 
visually clear. However, if we apply the 
extension, the algorithm correctly  determines that the two centers
are close, leading to a new round of optimization performed after
combining the two signal regions together. The resulting SNR for the
larger sets of ellipses is 36.91, large enough to claim that a flaw
has been detected.   
\subsection{Detection rule and statistical models for estimating POD}
\label{detection}
\subsubsection{Choice of a detection threshold}
\label{choice}
From the development in Section~\ref{algorithm}, we can
extract some important metrics from every indication (defined by the
inner and outer ellipses). These metrics -- the signal peak $\check Y$, 
noise peak $\check e$ and average noise $\bar e$ -- are  
representative of the image and the flaw and can be used to
calculate the SNR according to the definition in~Section~\ref{SNR}.  
Detection of a flaw is claimed whenever SNR $>$ $\alpha $, where
$\alpha $ is some SNR detection criterion that depends on the NDE
technology and the application. (For instance, $\alpha=2.5$ is
typically used in multi-zone ultrasonic inspection of titanium billets
and forgings.)  We make $\alpha $ adaptive to  our examples and 
facilitate comparisons by defining $\alpha $ as that 
value which gives an observed probability of false alarm (PFA) of some
pre-specified value (e.g. 3\% for the vibrothermography applications  
and $\le $ 1\% for the ultrasonic inspections), based on the images
corresponding  to specimens without any flaws. In this paper, we
follow \citet{olinandmeeker96}'s definition of the PFA as the
probability, for a particular inspection opportunity, of a flaw
determination when there is no flaw. Operationally, $\alpha $ is the
(1 -- PFA) quantile  of the SNR values for the observed noise images. 

Following~\citet{nietersetal95}, we define a noise threshold as $e_{th} 
=\alpha \check e + (1-\alpha)\bar e$, which 
is a random threshold that varies from one potential flaw to another
and that tends to be lower in regions with lower amounts of
background noise. The above detection criterion is then equivalent
to $\check Y>e_{th}$ (equivalently $D = \log_{10} \check Y - \log_{10}
e_{th} > 0$). Thus, we have an indication ({\em i.e.}, we claim a flaw
detection) whenever a specimen  
produces a positive value of $D$. Based on this criterion, each 
indication or candidate flaw in an  image can be classified as either
a detect (actual flaw) or a non-detect (noise artifact). In 
\begin{wrapfigure}{r}{0.5\textwidth}
\vspace{-0.25in}
\begin{center}
\mbox{
 \includegraphics*[width=0.49\textwidth]{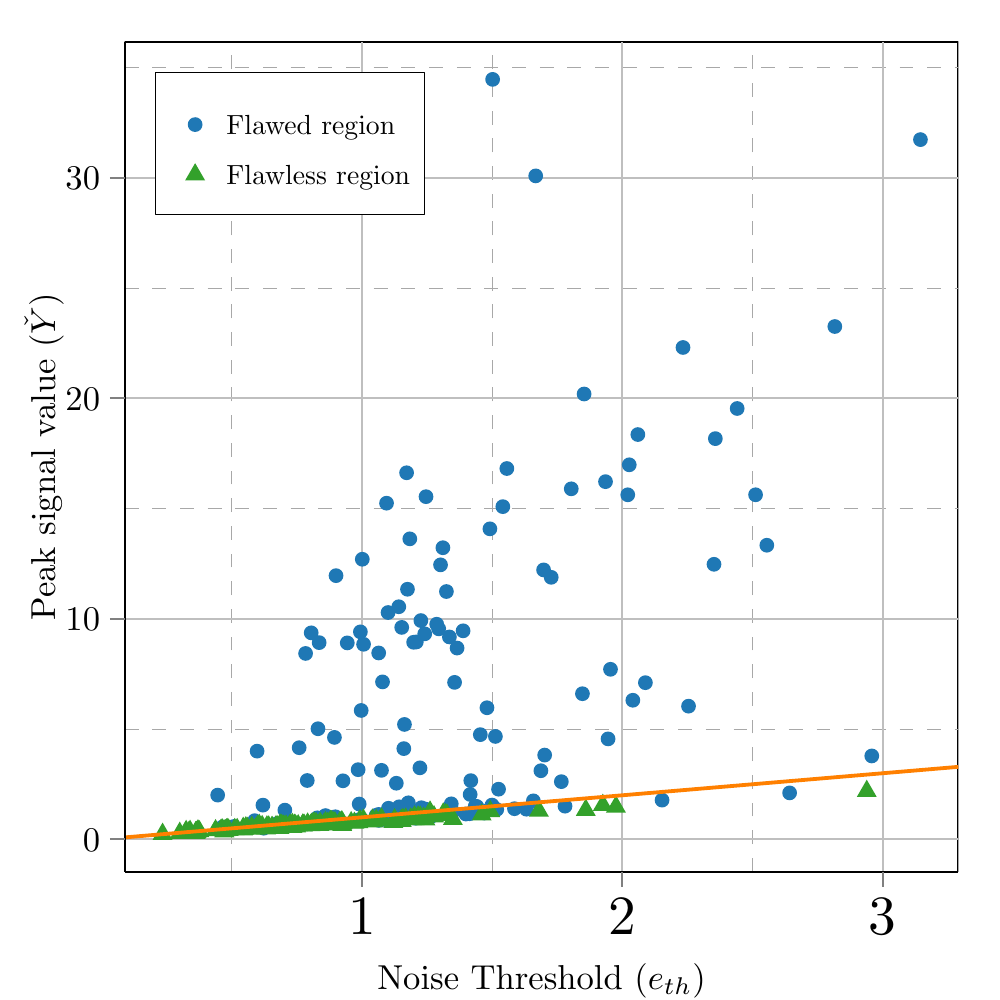}}
\end{center}
\vspace{-0.3in}
\caption{Classification of vibrothermography images, with detection
  claimed if $\check Y > {e}_{th}$, that is, if the  
data point is above the dashed line. Dots and triangles correspond to
flawed  and flawless regions respectively.}
\label{f6}
\vspace{-0.5in}
\end{wrapfigure}
our evaluations on vibrothermography noise images, we 
used $\alpha=2.354$, with  classification results as in Figure~\ref{f6}. 

We use statistical modeling to describe the relationship between
$D$ and flaw size. Figure~\ref{f7} is a scatterplot showing the
relationship between the 
observed metric $D$ and the logarithm (using
base 10, as is common in many engineering applications, including NDE)
of flaw size. The detection  criterion metrics  ($D$) for 
noise regions are also plotted in the figure and the 
horizontal dashed line represents the detection threshold, above which
one would  claim a flaw detection (a threshold of zero is used here).  

\subsubsection{The Noise-Interference Model}\label{NIM}
\begin{wrapfigure}{r}{0.5\textwidth}
\vspace{-0.525in}
\begin{center}
\mbox{
  \includegraphics*[width=0.49\textwidth]{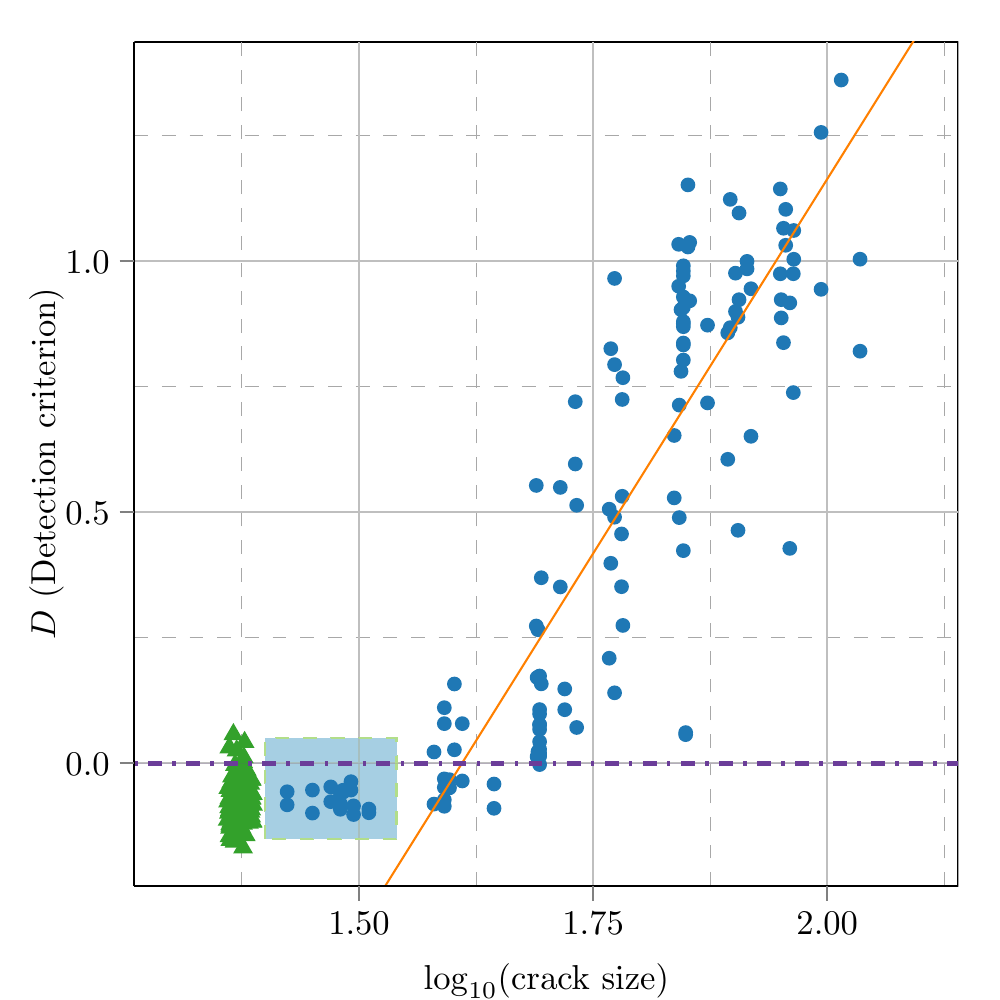}}
\end{center}
\vspace{-0.3in}
\caption{Plot  of $D$ (obtained after drawing the optimal
  ellipses 
  against  log${}_{10}$(flaw size) along  with the fitted regression line
  based on the   NIM model.}
\label{f7}
\vspace{-0.2in}
\end{wrapfigure}

Figure~\ref{f7} shows an approximately linear relationship
 between ${D}$ and $\log  
_{10} ({\mbox{crack size})}$ for 
 
\hspace{-0.31in} specimens with $\log _{10}
 (\mbox{crack size})$ larger than 1.56, ({\em i.e.}, signal data
 points that are  not in the highlighted rectangular region). This
  linear relationship levels off as $\log _{10} (\mbox{crack size})$
 falls below 1.55, corresponding to the $D$-values in the
 highlighted rectangle.  These $D$-metrics (in the highlighted
                                rectangular region) 
tend to have the same level as those for noise
specimens (represented by the $\blacktriangle$ symbols). This
leveling-off behavior can be explained by the fact that 
small flaws may be swamped by surrounding
   noise, so we conclude that the observed  response has a high
   probability of having been caused by a noise 
   artifact even in the presence of a small flaw. ~\citet{liandmeeker09}
   introduced the noise    interference model (NIM) 
   to describe this kind of relationship between a response and flaw
   size. According to this model for the vibothermography example
   (where the flaws are cracks), we consider the observed $D$-metric
   corresponding to a specimen with signal to be the competing result
   of signal ($D_{{\rm signal}} $) and noise ($D_{{\rm noise}} $),
   {\em i.e.}, $D_{{\rm obs}} =\max (D_{{\rm signal}} ,D_{{\rm
       noise}})$ where  $D_{{\rm signal}} =\beta _{0} +\beta _{1} \log
   _{10} (\mbox{crack size})+\varepsilon_{s}$,
where  $\varepsilon_{S} $ follows a normal distribution $N(0,{\rm \;
}\sigma _{S}^{2} )$ and $D_{{\rm noise}} \sim N(\mu _{N} ,{\rm \;
}\sigma _{N}^{2} )$. Based on this probabilistic model specification,
the observed likelihood for  specimens with a crack is given by
\begin{equation}
\begin{split}
L_{{\rm crack}} =\prod _{{\rm k\; with\; crack}} & {\sigma^{-1}_{S}}
  \phi\left[({D_{k} -\beta _{0} -\beta _{1} \times \log_{10}
      (\mbox{crack size})})/{\sigma _{S}} \right] \Phi\left[(D_{k} -\mu _{N})/\sigma _{N}
\right]+  \\
& \qquad \qquad
\sigma^{-1}_{N}  \Phi\left[\sigma_S^{-1}(D_{k} -\beta _{0} -\beta
      _{1} \log_{10} (\mbox{crack size}))\right]\phi
  \left[\sigma_N^{-1}(D_{k} -\mu _{N})\right] \\
\end{split}
\end{equation}
while the likelihood for flawless specimens is  
$L_{{\rm noise}} =\prod _{{\rm j\; without\; flaw}}{\sigma^{-1}_{N} }  \phi 
\left[\sigma_N^{-1}({D_{j} -\mu _{N} }) \right].$ 
Because specimens are independently inspected, the total likelihood
for all the specimens is, as usual, the product of the two likelihoods
$L=L_{{\rm crack}} L_{{\rm noise}}$, from which maximum likelihood
estimates (MLEs) of $\beta _{0} ,\beta _{1} ,\mu _{N} ,\sigma _{S} $, 
and $\sigma _{N} $ can be obtained in the usual way. The resulting  
fitted regression line for the NIM model after plugging in the MLEs
for our vibrothermography datasets  is shown in
Figure~\ref{f7}. Finally, the normality assumption in the NIM seems
reasonable (see Section~\ref{supp.diag} for details).

\section{Performance Evaluations}
\label{evaluations}
\subsection{Simulation Experiments}
We evaluate performance of our algorithm for 
automated optimal feature extraction as well as the use of the NIM on
the $D$-metrics obtained from the extracted features. Performance 
calibrations are in terms of the capability of our methods
and modeling to detect flaws of different sizes. 
\subsubsection{Performance Metrics}
\label{performance}
\begin{wrapfigure}{r}{0.5\textwidth}
\vspace{-0.1in}
\mbox{
  \subfigure[]{\includegraphics[width=0.25\textwidth]{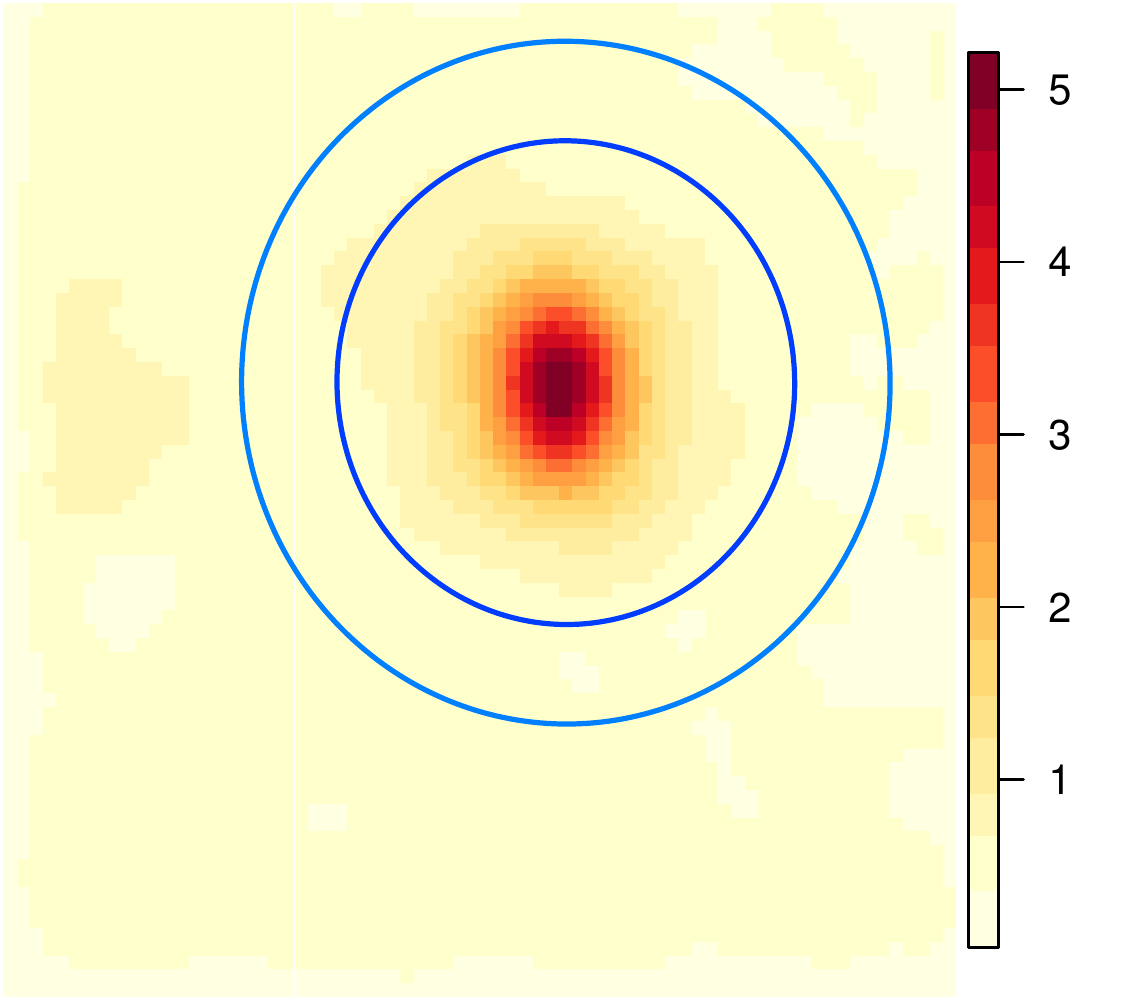}}
  \subfigure[]{\includegraphics[width=0.25\textwidth]{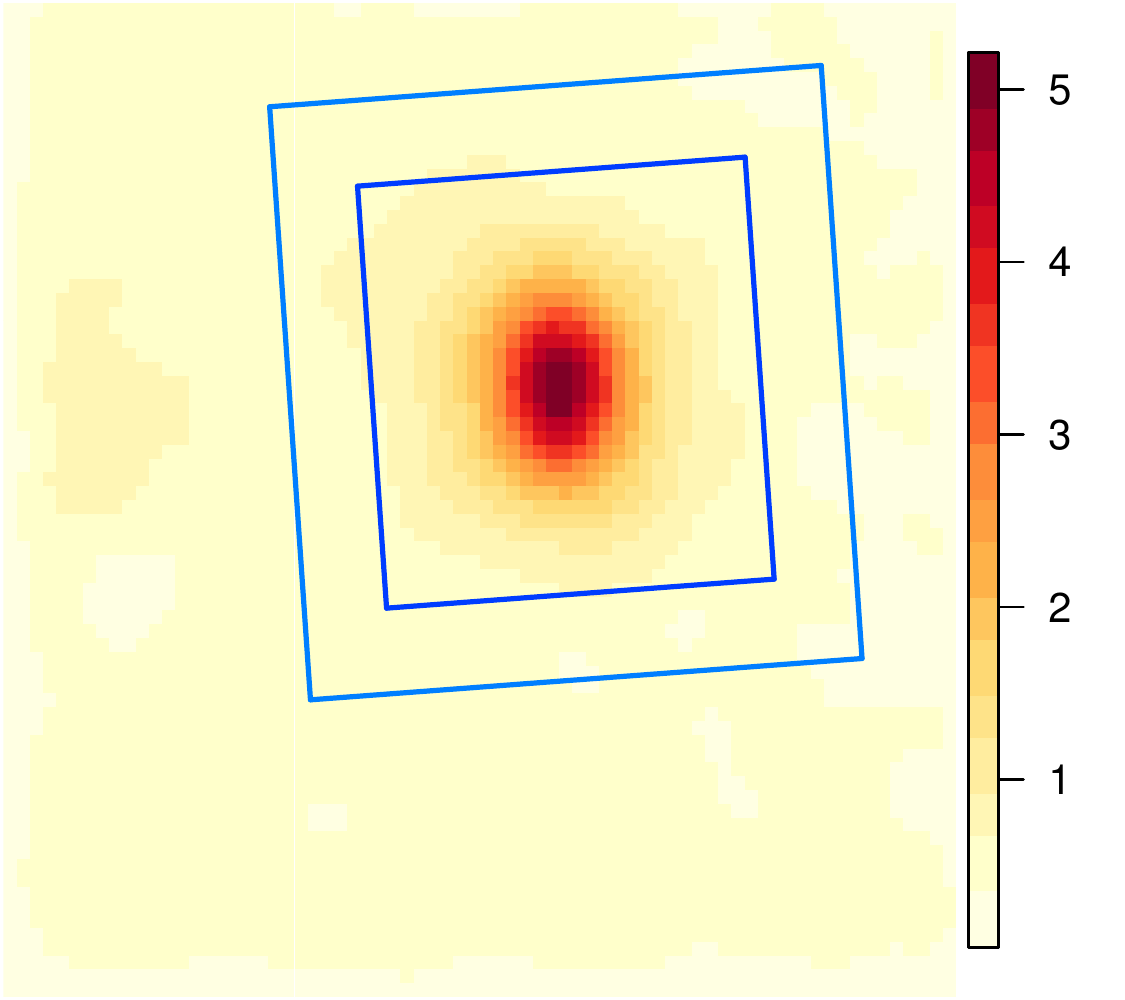}}
}
\vspace{-0.1in}
\caption{Results of using the algorithm on feature extraction using
  the proposed algorithm with (a) drawn optimal inner and outer
  ellipses and (b) drawn optimal optimal inner and outer rectangles.} 
\label{f9}
\vspace{-0.2in}
\end{wrapfigure}

We follow the NDE practice of quantifying detection capability in terms of the
POD which is also the most commonly-used metric in the NDE and
statistical literature. According to the NIM model of 
Section~\ref{NIM}, we can (see the derivations in
Section \ref{supp.deriv}) express the POD  as a function of flaw size 
and the regression parameters using the model:
\begin{equation}
\mbox{POD(flaw)} = \Pr(D_{obs} > 0)  =1-{\rm \Phi }\left[
-\frac{\beta _{0} +\beta _{1} \log_{10} ({\rm flaw})}{\sigma _{S} 
} \right]\Phi \left(-\frac{\mu _{N} }{\sigma _{N} } \right).
\label{ZEqnNum709216}
\end{equation}
This definition of POD takes credit for a detection for cases where a
noise artifact results in a signal stronger than that from the actual
flaw and results in an above-threshold observed signal.
Our performance evaluations compare our proposed  algorithm with
alternatives in terms of the POD. These alternatives
employed (a) the commonly-used ``peak amplitude'' method described in
Section~\ref{background} and (b) an automated and optimized version of
the~\citet{howardandgilmore94} rectangle-drawing method which uses a
development parallel to Section~\ref{optimalelliptical} except that it
is used to draw an optimal  inner rectangle. We call this method
``Rectangle'' and  note that it is expected to be as good, if not
better, than an operator-drawn rectangular set (which is subjective
and does not involve optimality 
considerations). For easy reference, we use ``Ellipse'' to refer to
our proposed method based on drawing optimal 
inner and outer ellipses. Figure~\ref{f9} displays a comparison of the
results obtained using inner and outer 
ellipses (left) and inner and outer rectangles (right). 
Similar to the developments in Section~\ref{optimalelliptical}, the
``Rectangle'' method is also readily extended 
for multiple targets and single targets with paired hotspots. 
The POD impression for the ``Rectangle'' method is the same as
that for our ``Ellipse'' method and is given
by~\eqref{ZEqnNum709216}. The expression is, however, not applicable to
the ``peak amplitude'' method (henceforth abbreviated as ``PeakAmp'')
so we develop the POD for this case next. 

The ``PeakAmp'' method uses the peak intensity $\check Z$ {\em of
  the raw unprocessed hottest frame image} to characterize  the
image. Here we have a signal pixel $Y_S$ related to the flaw size
by the model $\log_{10} Y_S = \gamma_0 +\gamma_1  \log_{10}
(\mbox{flaw}) + \upsilon_S$, where the parameter $\gamma_0$
controls the magnitude of the hottest pixel, $\gamma_1$ reflects
the relationship between peak amplitude and flaw size, and $\upsilon_S\sim 
N(0,\kappa^2_S)$. For a noise pixel, we have $\log_{10} Y_N\sim N(\nu_N,
\kappa^2_N)$. This relationship was inspired by the finding 
in many of our (real-life) datasets that the logarithm of peak
amplitude and the logarithm of flaw size is approximately
linearly-related. For some inspection
methods, including ultrasound, there is a physics-based explanation
for this relationship, as described in~\citet{lietal14}.
Using the NIM model  to describe the relationship between 
the model response ({\em i.e.}, $\log _{10} \check Z$) and 
flaw size, the POD \citep[for detailed
  derivation, see Section~\ref{supp.deriv} or Equation~(3)
  of][]{liandmeeker09} is 
\begin{equation*}
{\rm POD}\left({\rm flaw}\right)  = \Pr \left(\check Z >Z_{{\rm th}} \right) \\
 =1-{\rm 
\Phi }\left(\frac{\log_{10}(Z_{\rm th}) - \gamma _{0} - \gamma _{1} {\rm log}_{10} \left({\rm flaw}\right)}{\kappa _{S} } \right)\Phi \left(\frac{\log _{10}(Z_{{\rm th}}) -\nu _{N} }{\kappa
_{N} } \right), 
\end{equation*}
where $Z_{{\rm th}}$ is the detection threshold for the peak-amplitude
response and $(\gamma_0,\gamma_1,\kappa_S,\nu_N,\kappa_N)$ are estimated
using the MLEs in the same manner as before. 

\subsubsection{Results}
\begin{wrapfigure}{r}{0.5\textwidth}
\vspace{-0.2in}
\begin{center}
\mbox{
  \includegraphics*[width=0.5\textwidth]{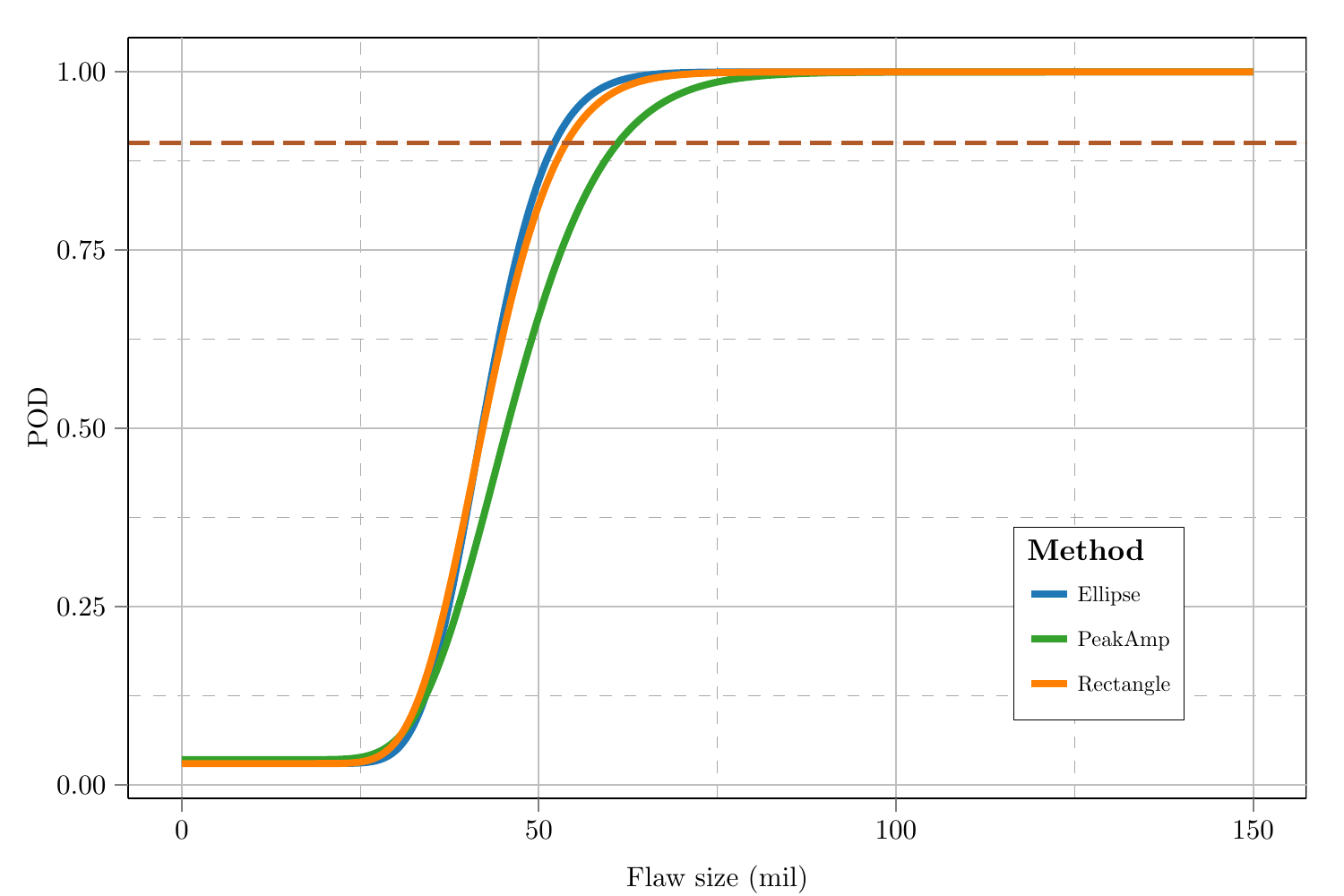}
}
\vspace{-0.4in}
\caption{POD curves for the proposed algorithm and the two alternative methods.}
\label{f10}
\end{center}
\vspace{-0.4in}
\end{wrapfigure}
Figure~\ref{f10} compares the POD curves corresponding to the
``Ellipse'' algorithm  with those  obtained 
using the ``Rectangle'' and ``PeakAmp'' methods. The
PFA-values for all methods have been calibrated to be the same (0.03 here) to
allow a fair comparison. The flaw size at which each curve attains a
POD of 0.9 (indicated by the horizontal dashed line) is called the 
$a_{90} $ value in NDE parlance. In the NDE community, the estimated
$a_{90}$  is widely used as a scalar metric of inspection
capability and to 
compare  different NDE methods. A smaller $a_{90} $ value is usually a
sign of a better, more sensitive, procedure.  Figure~\ref{f10}
thus shows that both the SNR-based ``Ellipse'' and
``Rectangle'' algorithms outperform the simpler ``PeakAmp'' method
by yielding better POD curves and  smaller $a_{90} $ values. Further,
 ``Ellipse''  has somewhat higher POD values
than ``Rectangle'' for  almost the entire range of the flaw sizes
in consideration. The $a_{90} $ estimates for the  ``Ellipse'',
``Rectangle'' and ``PeakAmp'' methods are 52.2 mils, 53.8
mils and 61.0 mils, respectively. These $a_{90} $ values 
suggest  that the proposed ``Ellipse'' method performs better
than the two  
other methods, indicating significant improvement over the alternatives.  
Note that the most important part of the POD curve, from a practical
perspective, is where probability of detection is relatively high. A
common saying in the NDE  field is ``It is not the smallest flaw we
might detect that is of interest, but rather  the largest flaw we
might miss.'' Under this adage,  ``Ellipse''  is a better performer
than the ``Rectangle'' or ``PeakAmp'' methods. 

We also evaluated the performance of our algorithm beyond the original
datasets on a series of simulated NDE images. The simulated images are
composed of two parts: the signal and the noise. The noise images for our
simulations were obtained by resampling (with replacement) from the
pool of flawless vibrothermography images (also called noise
images). For our simulated 
flawed images, we added -- to these resampled noise images -- signal
images from a Gaussian signature  with  peak amplitude defined as
before, that is, $\log_{10} (\mbox{peak amplitude})=\gamma _{0}
+\gamma _{1}  \log_{10} (\mbox{flaw})$.
\begin{figure}[h]
\begin{center}
\mbox{
\subfigure[]{\includegraphics*[width=0.75\textwidth]{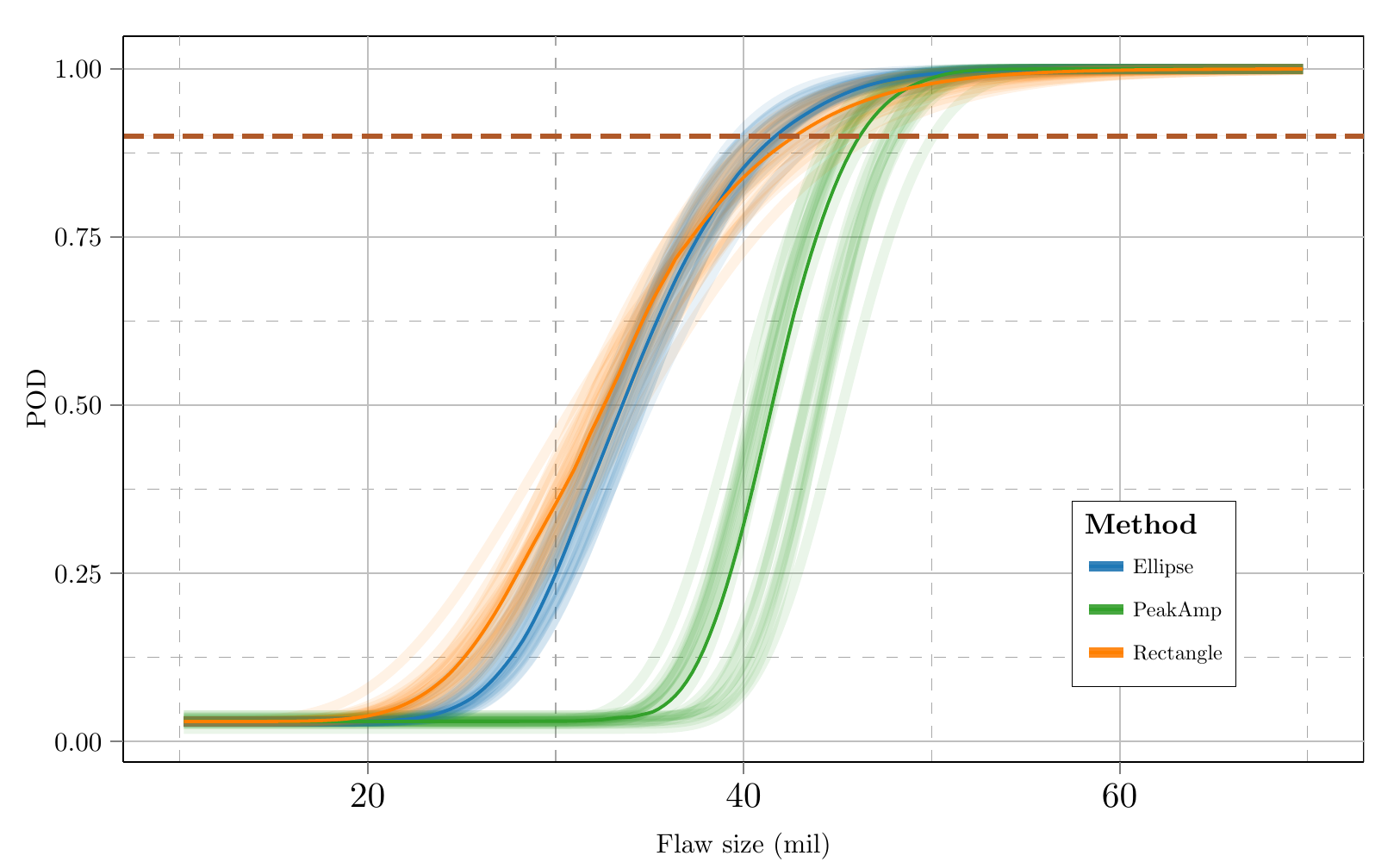}}
\subfigure[]{\includegraphics*[width=0.25\textwidth]{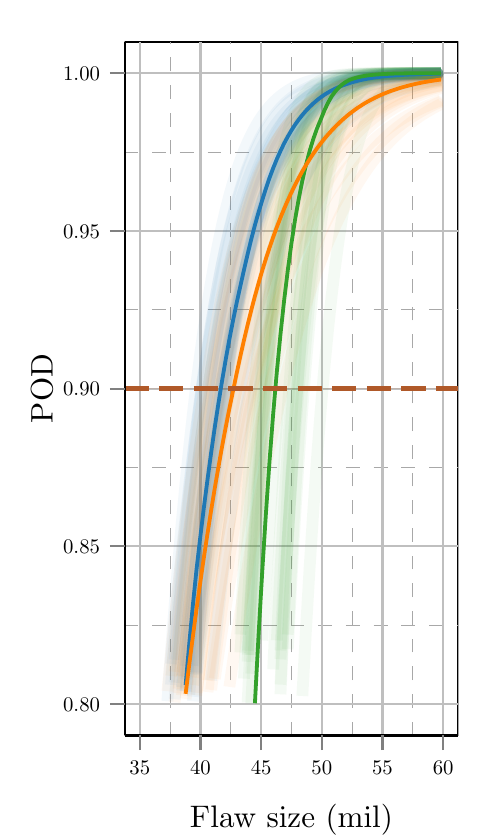}}
}
\end{center}
\vspace{-0.2in}
\caption{(a) POD comparison of three methods based on 20
  simulations and (b) a closer view of the POD comparison for flaw
  sizes between 35 and 60 mils.}
\label{f1213}
\vspace{-0,1in}
\end{figure}
The FWHM $\hbar$ of the 
Gaussian signature is proportional to the flaw size 
where the constant of this proportionality is given by $k$. In our
simulations, we used $\gamma _{0} =-7$, $\gamma _{1} =1.22$ and
$k=0.0785$. For each of the flaw sizes for the specimens used in the
vibrothermography inspection, we obtained 20 simulated degraded signal images
and applied the three detection  methods to them as well as to the
noise images. Thus, three POD curves were produced and corresponding
$a_{90} $ values were computed to compare the three methods. 
The resulting 20 sets of POD curves are shown in
Figure~\ref{f1213}. Clearly, the ``Ellipse'' method  generally
produces better POD curves than the ``Rectangle'' method 
but both approaches produce better POD curves than the
``PeakAmp'' methods.

\begin{wrapfigure}{r}{0.5\textwidth}
\begin{center}
\vspace{-0.65in}
\mbox{
 \includegraphics*[width=0.5\textwidth]{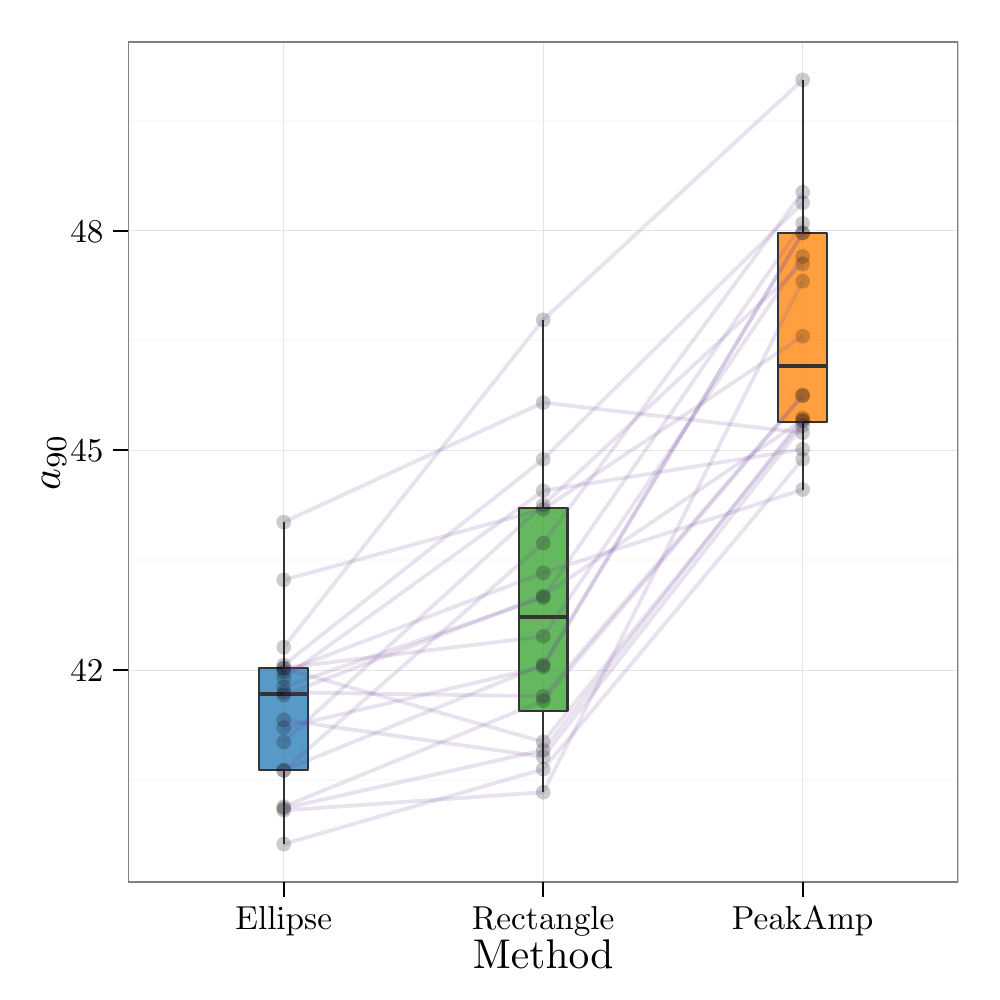}}
\end{center}
\vspace{-0.2in}
\caption{Estimates of $a_{90}$ obtained by the three
  methods. The parallel coordinates plot
  connects the estimates obtained for the three methods from each experiment.}
\label{1415}
\vspace{-0.3in}
\end{wrapfigure}
Figure~\ref{1415}  displays  estimated $a_{90}$ values
obtained from the three methods. We see relatively good separation
among the three $a_{90} $ distributions. Also,  ``Ellipse'' tends to
produce  the smallest $a_{90} $ values. Thus, in general, it performs better
than ``Rectangle'' which in turn 
performs better than ``PeakAmp''. The parallel
coordinates plot underlying the boxplots indicates better performance
of ``Ellipse'' over ``Rectangle'' for all but two cases where
``Rectangle'' performs slightly better. Both of these methods are almost
always better than ``PeakAmp'' (with the ``Rectangle'' 
method being
slightly outperformed in only one case, and  
``Ellipse'' being
always superior). This assertion was confirmed by the results of a
paired Wilcoxon signed rank 
test between the $a_{90}$-values obtained using ``Ellipse'' and
``Rectangle'' (which produced  a $p$-value of 0.0002 and a similar
test between the $a_{90}$-values obtained using ``Rectangle'' and ``PeakAmp'' $a_{90} $-values ($p$-value=$3.82\times 10^{-6}$). 
These simulation results are consistent with what we observed 
in the real vibrothermography datasets. Taking into consideration both
the real vibrothermography  
data analysis and the simulation results, we are confident that the
proposed ``Ellipse'' method is a strong candidate for analyzing
NDE image data, in the sense of producing  high POD. We now analyze
the dataset introduced in Section~\ref{vbexample}. 

\subsection{Application to Vibrothermography Samples}
\label{application}
The three raw images introduced in Figure~\ref{fig1} with
postprocessed (matched filtered) versions displayed in Figure
\ref{fig2} were analyzed using the methods illustrated in this
paper. Under the ``Ellipse'' method, the SNR  for the image with the
strong signal (Figure~\ref{fig1}a) was computed to be 16.12, while
that for the weak 
signal (Figure~\ref{fig1}b) was 5.56 and the noise image
(Figure~\ref{fig1}c) of no true flaw was calculated to be 2.30. 
No matter whether we use the threshold of
$\alpha= 2.354$ as obtained using the development in
Section~\ref{choice} above, or the industry standard of 2.5, the first
two images are correctly classified as signal images  (which means
that they contain flaws, which were cracks in this application) and the
last image is correctly classified as solely containing
noise (or image of a flawless specimen). Under the ``Rectangle'' method,
the SNRs were computed to be 14.52, 5.77 and 2.37 for the three images
in Figure~\ref{fig1}a, \ref{fig2}b and \ref{fig1}c,
respectively. Using developments similar to 
Section~\ref{choice}, but for the ``Rectangle'' method, we got the
cut-off of 2.5 (which is the same as the industry standard threshold) and
therefore the conclusions for these three specimens match the ones
for the ``Ellipse'' method. We stress here that the automated and
optimized version that is our ``Rectangle'' method, and which by
itself is a contribution of this paper, is not actually
used in NDE; current NDE practice utilizes
subjective operator-assisted rectangle-drawing. Using  the ``PeakAmp''
method however, 
the values obtained (of the pixel with the highest raw intensity) are
0.35, 0.09 and 0.053 for the three images in Figures
\ref{fig1}a, b, and c. The cut-off value for the ``PeakAmp'' method to
separate flaw from noise was derived using the methods in
Section~\ref{choice} as 0.104, so only the image with the strong
signal is identified as such and the method is unable to
distinguish between a weaker signal and noise, classifying both as
noise specimens. Our results  provide confidence in the
applicability of the ``Ellipse'' (and ``Rectangle'') methodology for
automated flaw detection in 
vibrothermography images. We note also that the SNR for the first
image is much larger than that for the second image, a finding that is
in good agreement with the fact that the first specimen has a larger
crack than  the second specimen.

\section{Discussion}
\label{discussion}
In this paper we developed an automated flaw detection algorithm based
on image processing and SNR detection. By setting the same probability
of false alarm  (PFA), the proposed automated algorithm based on
drawing (optimal) ellipses around the signal shows better  detection
performance than the alternative that is based on a simplistic
peak-amplitude scalar response or an intermediate algorithm that is an
automated and optimized version of a method where the operator draws
rectangles based on visual inspection. The POD  values tend to be
higher  for the proposed algorithm than the other two methods for the
flaw size range of  interest. Correspondingly, the $a_{90}$ value for
the proposed algorithm is smaller than that of the other two
algorithms. The simulation results based on simulating Gaussian-like
signal and resampling of the vibrothermography noise images confirm  
that the proposed algorithm outperforms the other algorithms. 

There are a number of areas that would benefit from increased
attention. Our methods here  followed the industry standard in
adopting the hottest frame for analysis. It would be interesting to
see if some other approach can provide similar or better results. We have
also here demonstrated and evaluated performance in terms of the
analysis of vibrothermography (and while not presented in detail in
this paper, ultrasound) images. It would also be important  to
see if our methodology can be adapted and extended to other NDE
imaging inspection methods such as eddy current or X-ray
inspection. We expect that the answer is in the affirmative, but this
will need to be validated and tested. Finally, we hope that our
results will also spur greater statistical interest and involvement in
this important  industrial field of application.
\section*{Acknowledgements}
\if0\blinded {
This work was performed with partial support from the Federal
Aviation Administration under contract number  DTFACT-09-C-00006 through the
Center for Nondestructive Evaluation at Iowa State University. 
The ultrasonic test images used in Section~\ref{extensions} were
acquired as part of a project to study the POD of ultrasonic
inspections of forgings, supported  by  the Federal Aviation
Administration under contract number 08-C-00005 to Iowa State
University. We thank Tim Gray for providing the images
and for helping us to interpret them.
We also thank the Editor, an
Associate Editor and reviewers for suggestions that helped us to
improve an earlier version of this paper. 
}
\fi
\if1\blinded {
We thank the Editor, an
Associate Editor and reviewers for suggestions that helped us to
improve an earlier version of this paper. 
}
\fi


\renewcommand{\baselinestretch}{1}\normalsize
\begingroup
\def\bibfont{\small}
\bibliographystyle{asa}
\addcontentsline{toc}{section}{References}
\bibliography{refs}
\endgroup

\end{document}



\if0\blinded
{
\markboth{Ye et al{\it et al.}}{Automated Flaw Detection in NDE
  Images}
}
\fi
\if1\blinded
{
\markboth{}{Supplement to Automated Flaw Detection in NDE Images}
}
\fi

\title{\vspace{-0.8in} Supplement to ``A Statistical Framework for
  Improved Automatic Flaw Detection in Nondestructive Evaluation Images''} 
\if0\blinded
{
\author{
  \vspace{-0.1in}
  {\bf Ye Tian}\\
  \vspace{-0.2in}
  Department of Statistics and Statistical Laboratory\\
    Iowa State University\\
    Ames, IA 50011\\
    (tianye1984@gmail.com)\\
\and
    \vspace{-0.2in}
    {\bf Ranjan Maitra}\\
    \vspace{-0.2in}
    Department of Statistics and Statistical Laboratory\\
    Iowa State University\\
    Ames, IA 50011\\
    (maitra@iastate.edu)
    \and
    \vspace{-0.2in}
           {\bf William Q. Meeker}\\
           \vspace{-0.2in}
           Department of Statistics and the Center for Nondestructive
           Evaluation\\
           Iowa State University\\
           Ames, IA 50011\\
           (wqmeeker@iastate.edu)
           \and
           \vspace{-0.2in}
                  {\bf Stephen D. Holland}\\
                  \vspace{-0.2in}
                  Department of Aerospace Engineering and Center for Nondestructive Evaluation\\
                  Iowa State University\\
                  Ames, IA 50011\\
                  (sdh4@iastate.edu)
}
\date{\vspace{-0.5in}}
}
\fi
\if1\blinded
    {
      \date{\vspace{-1.2in}}
      \renewcommand{\baselinestretch}{1.4}\normalsize
    }
\fi
\maketitle
\renewcommand{\baselinestretch}{1.4}\normalsize

\renewcommand\thefigure{S-\arabic{figure}}
\renewcommand\thetable{S-\arabic{table}}
\renewcommand\thesection{S-\arabic{section}}
\renewcommand\thesubsection{S-\arabic{section}.\arabic{subsection}}
\renewcommand\theequation{S-\arabic{equation}}
\section{Notations used in Supplement}
In this supplement, references to sections, figures and equations in
the main paper are referred to using the same identifiers as in the
main paper. References to sections, figures and equations in the
supplement use the suffix ``S-''. 
\section{Methodology -- Supplement}
\subsection{Illustrative Examples showing the Effect of $\lambda$}
\label{S-lambda}
\begin{figure}[!h]
\vspace{-0in}
\begin{center}
\mbox{
\subfigure[Strong signal,
  $\lambda=2$]{\includegraphics[width=0.33\textwidth]{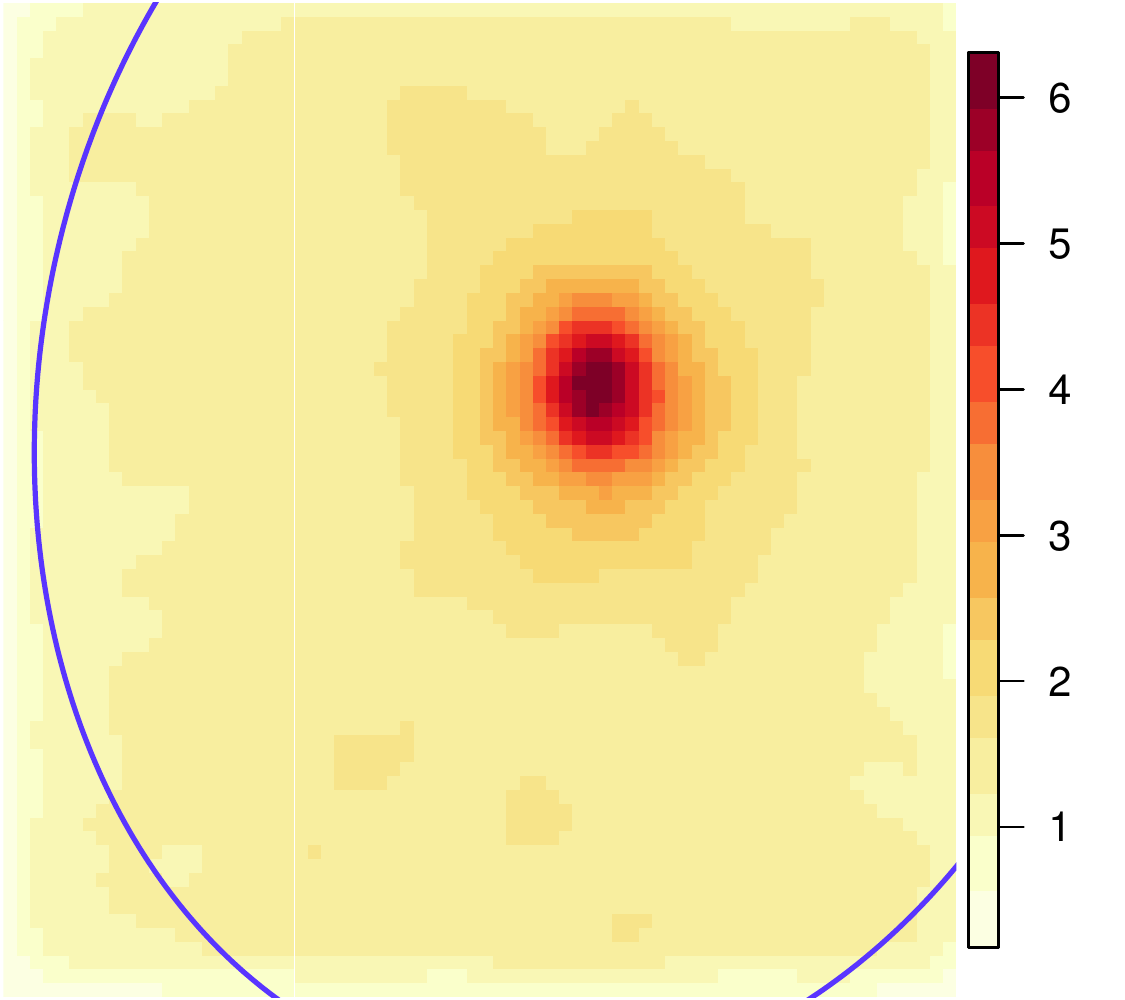}}
\subfigure[Strong signal,
  $\lambda=100$]{\includegraphics[width=0.33\textwidth]{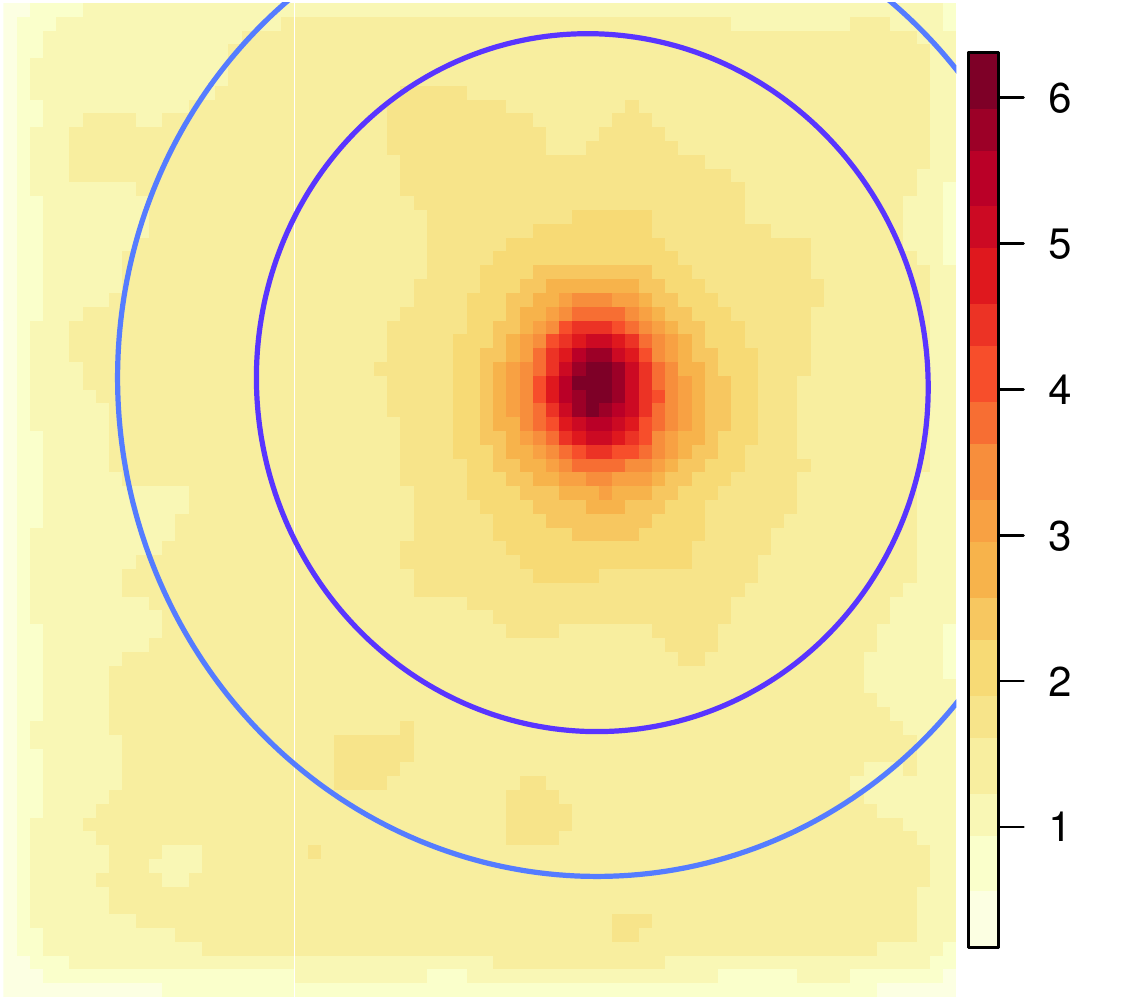}}
\subfigure[Strong signal,
  $\lambda=200$]{\includegraphics[width=0.33\textwidth]{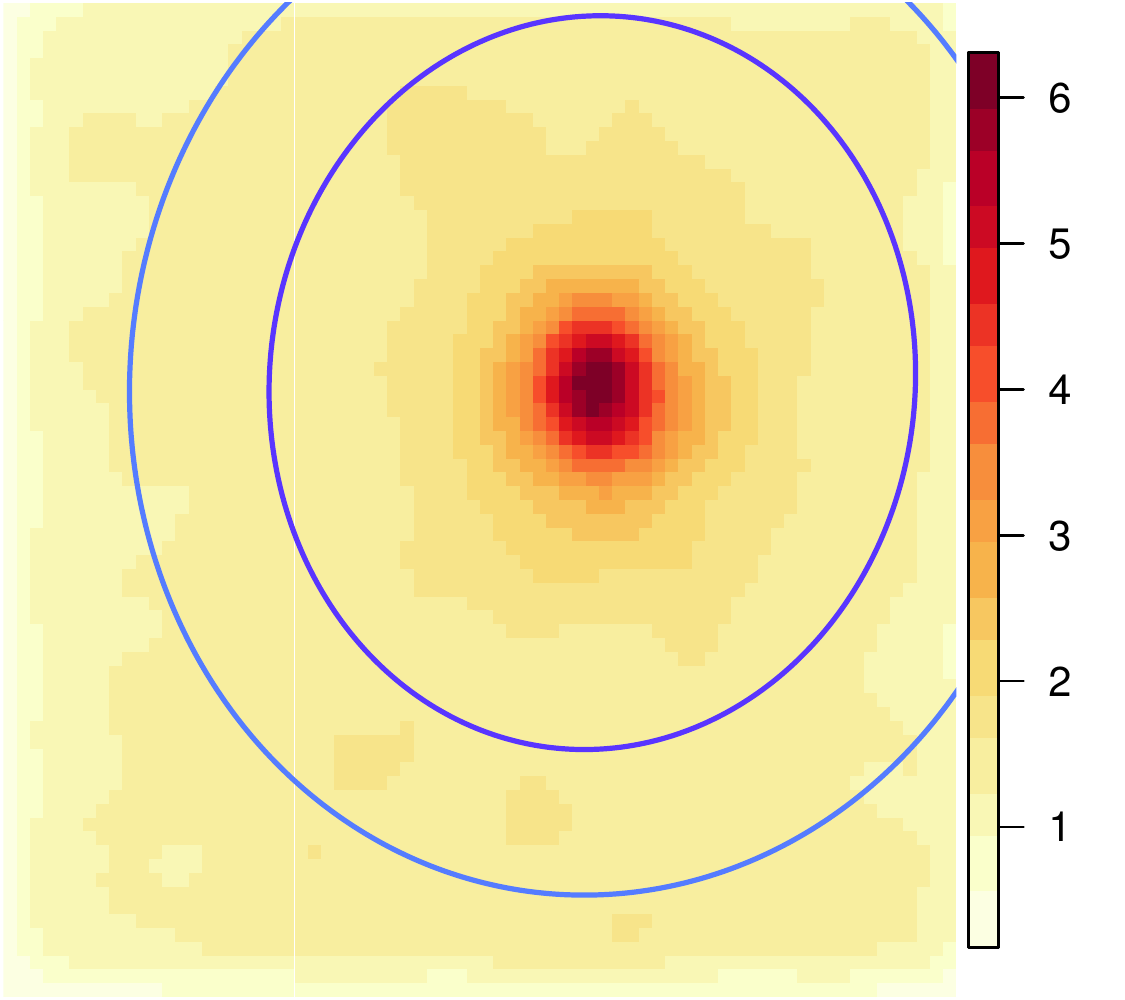}}
}
\mbox{
\subfigure[Weak signal,
  $\lambda=2$]{\includegraphics[width=0.33\textwidth]{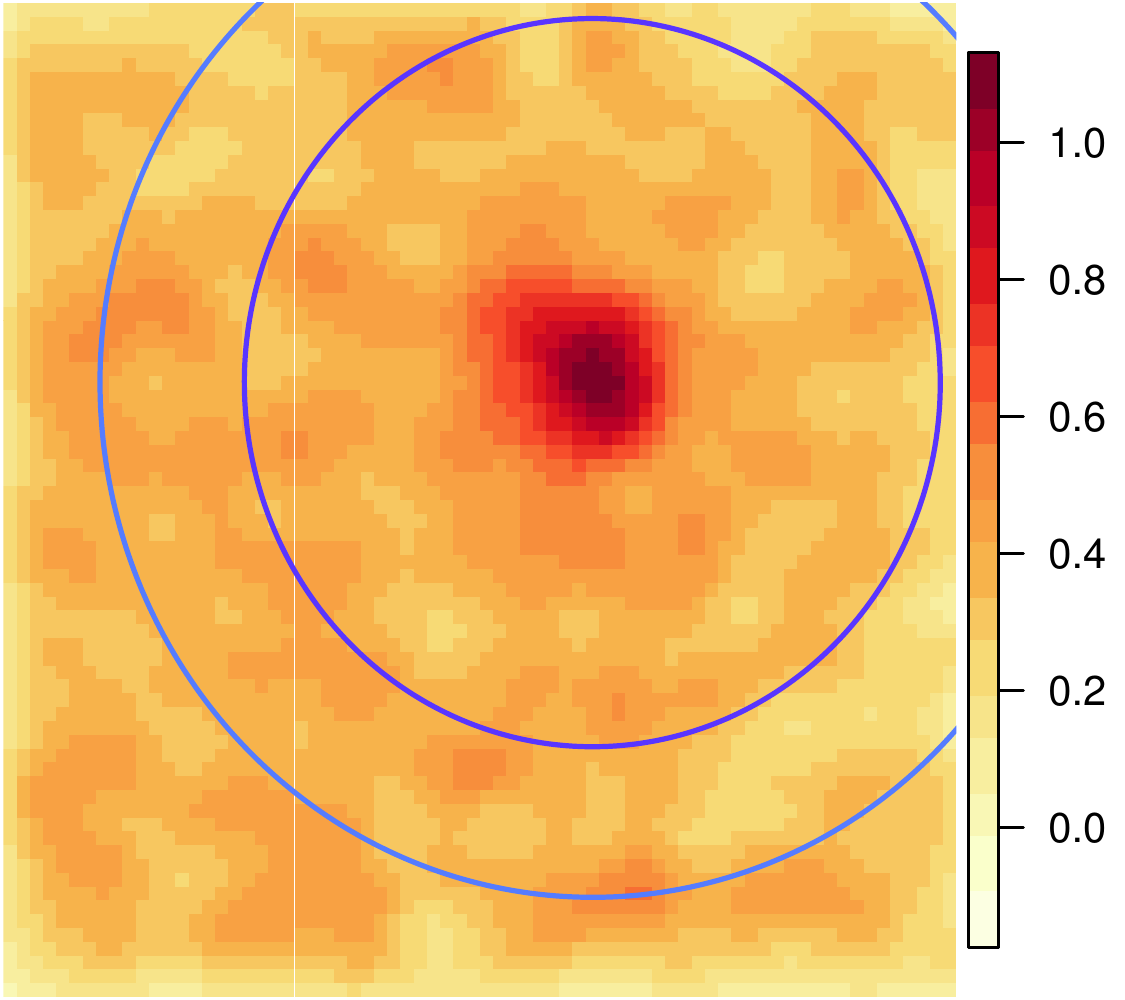}}
\subfigure[Weak signal,
  $\lambda=100$]{\includegraphics[width=0.33\textwidth]{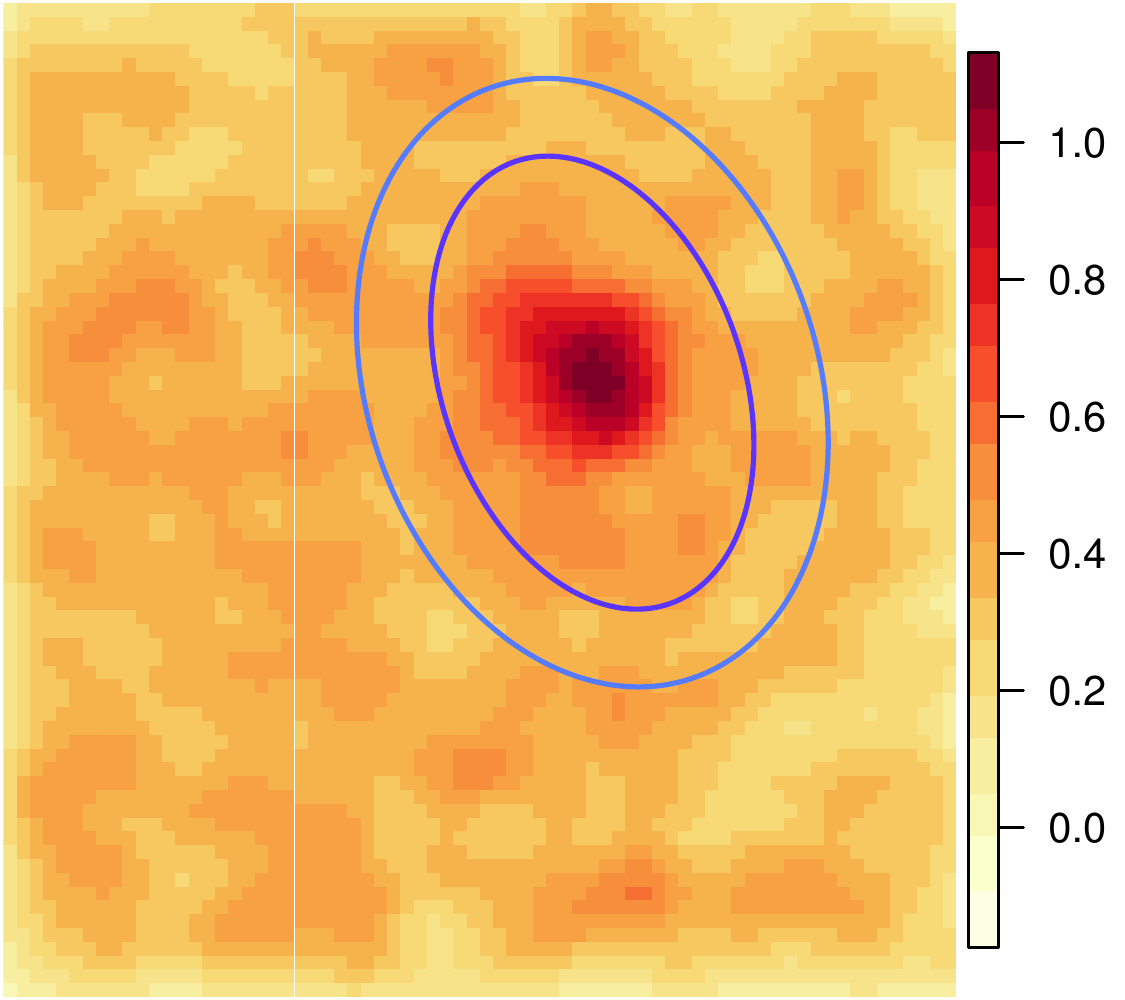}}
\subfigure[Weak signal,
  $\lambda=200$]{\includegraphics[width=0.33\textwidth]{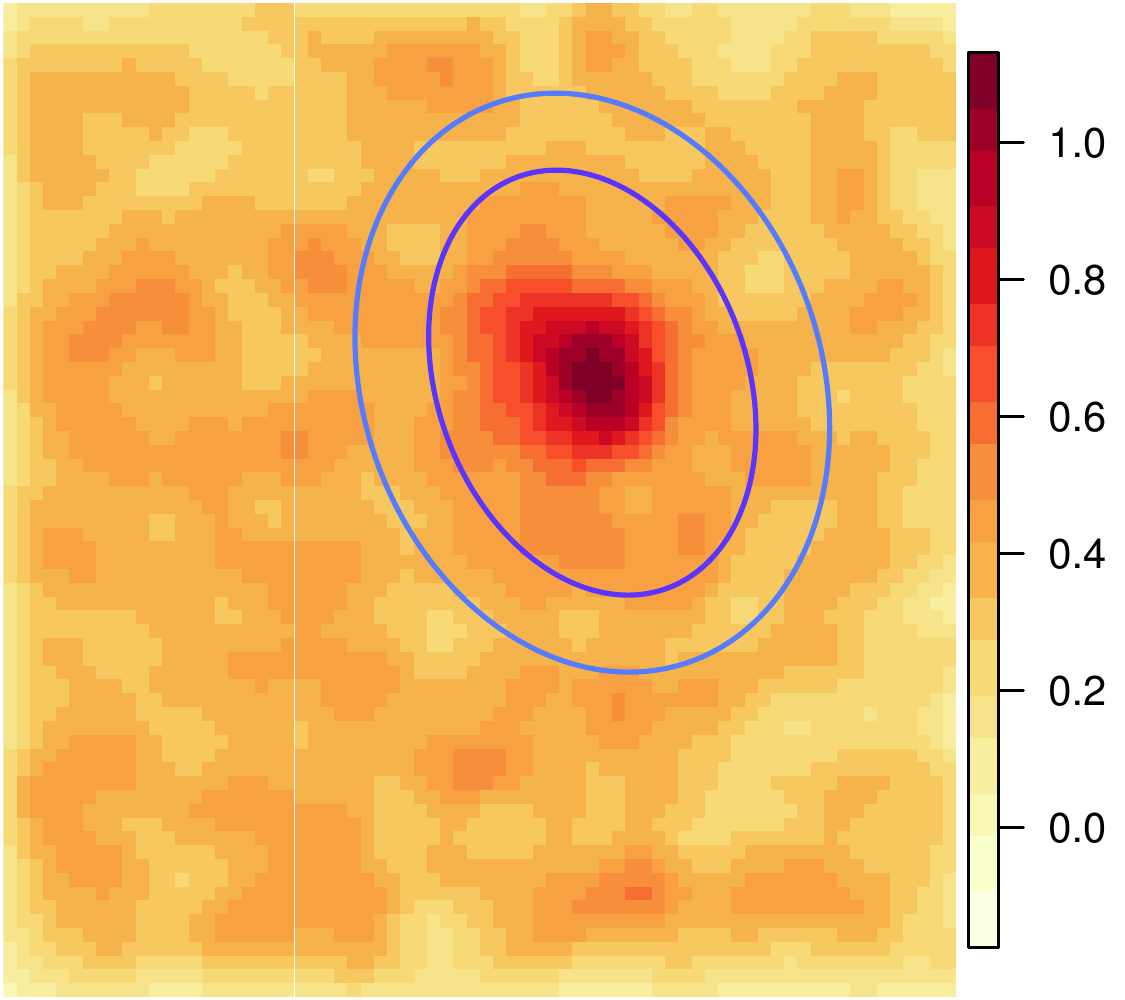}}}
\mbox{
\subfigure[Noise,
  $\lambda=2$]{\includegraphics[width=0.33\textwidth]{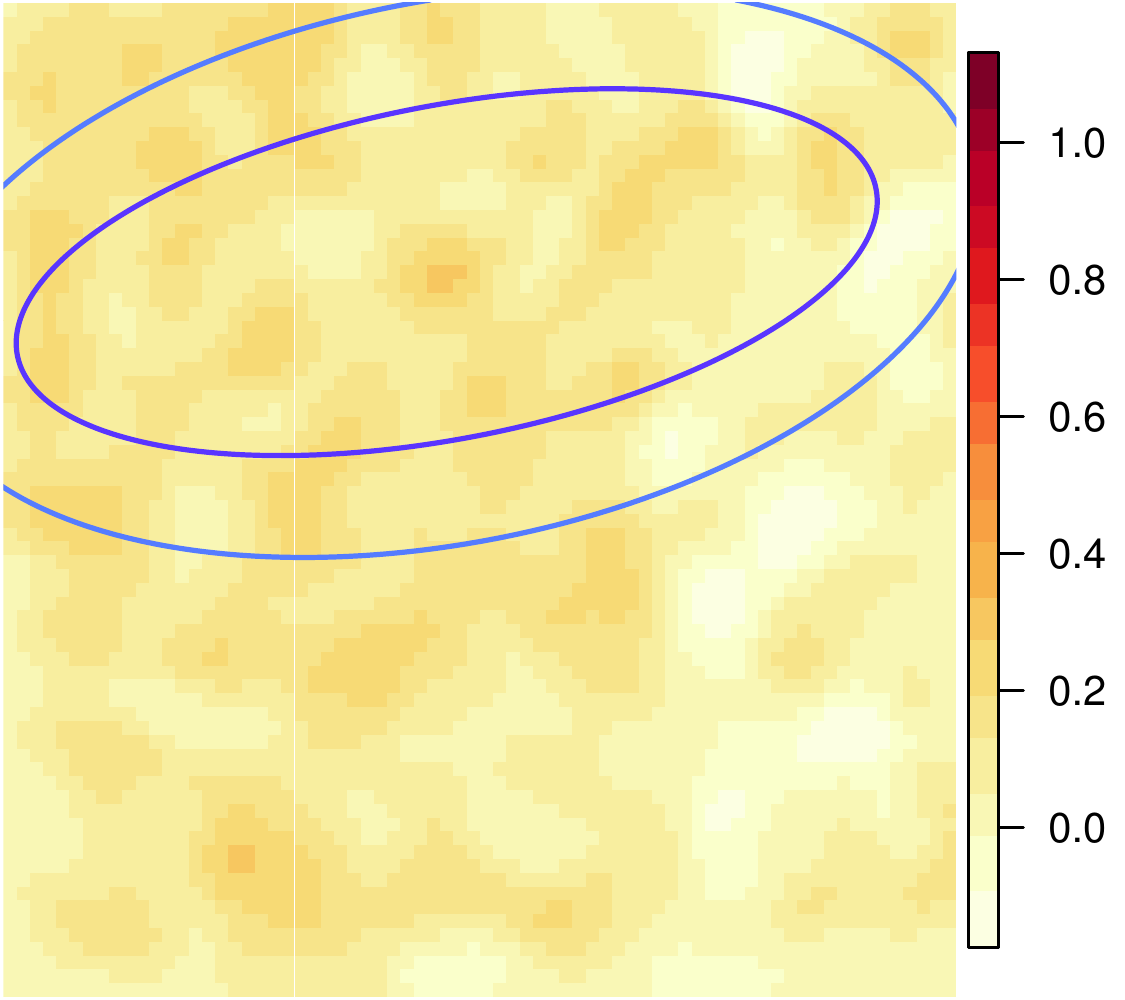}}
\subfigure[Noise,
  $\lambda=100$]{\includegraphics[width=0.33\textwidth]{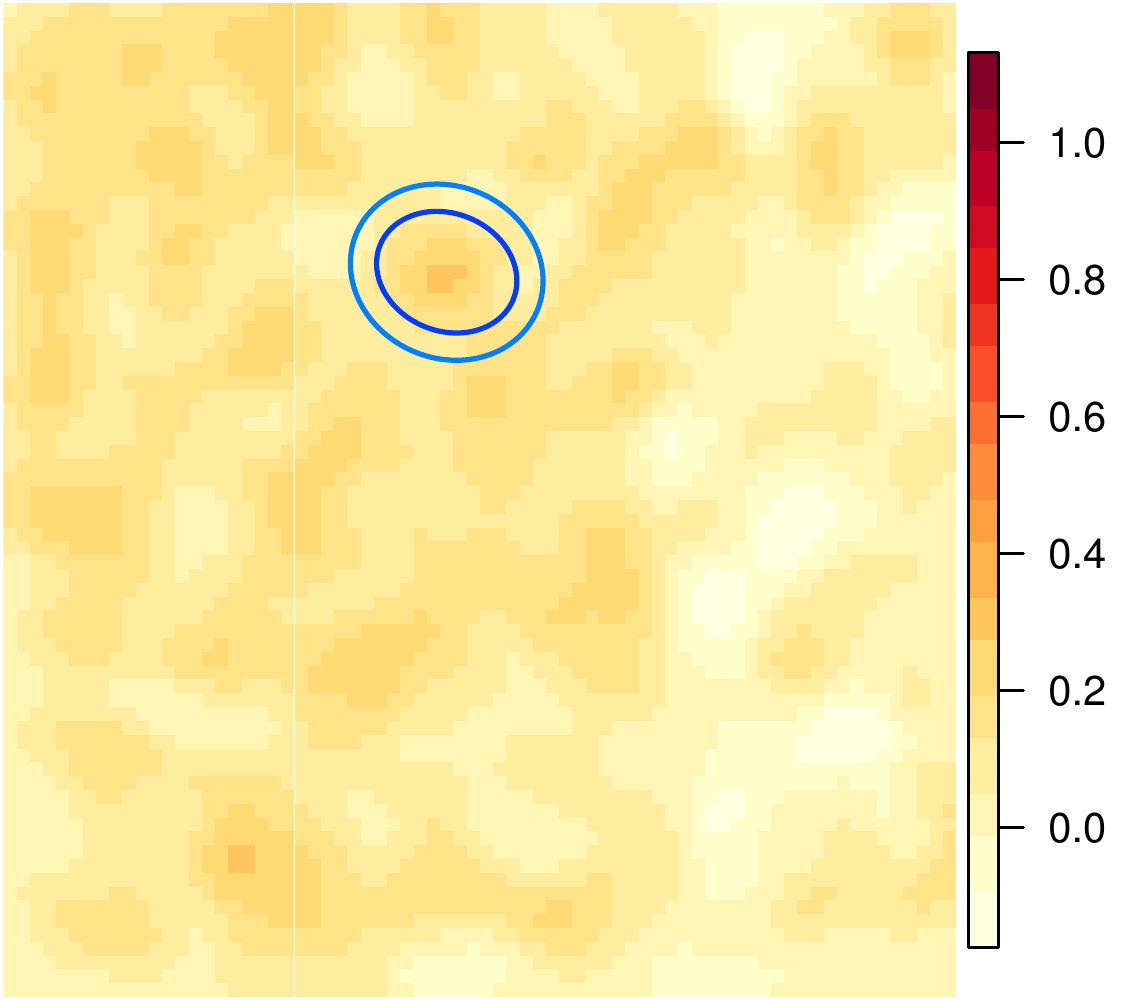}}
\subfigure[Noise,
  $\lambda=200$]{\includegraphics[width=0.33\textwidth]{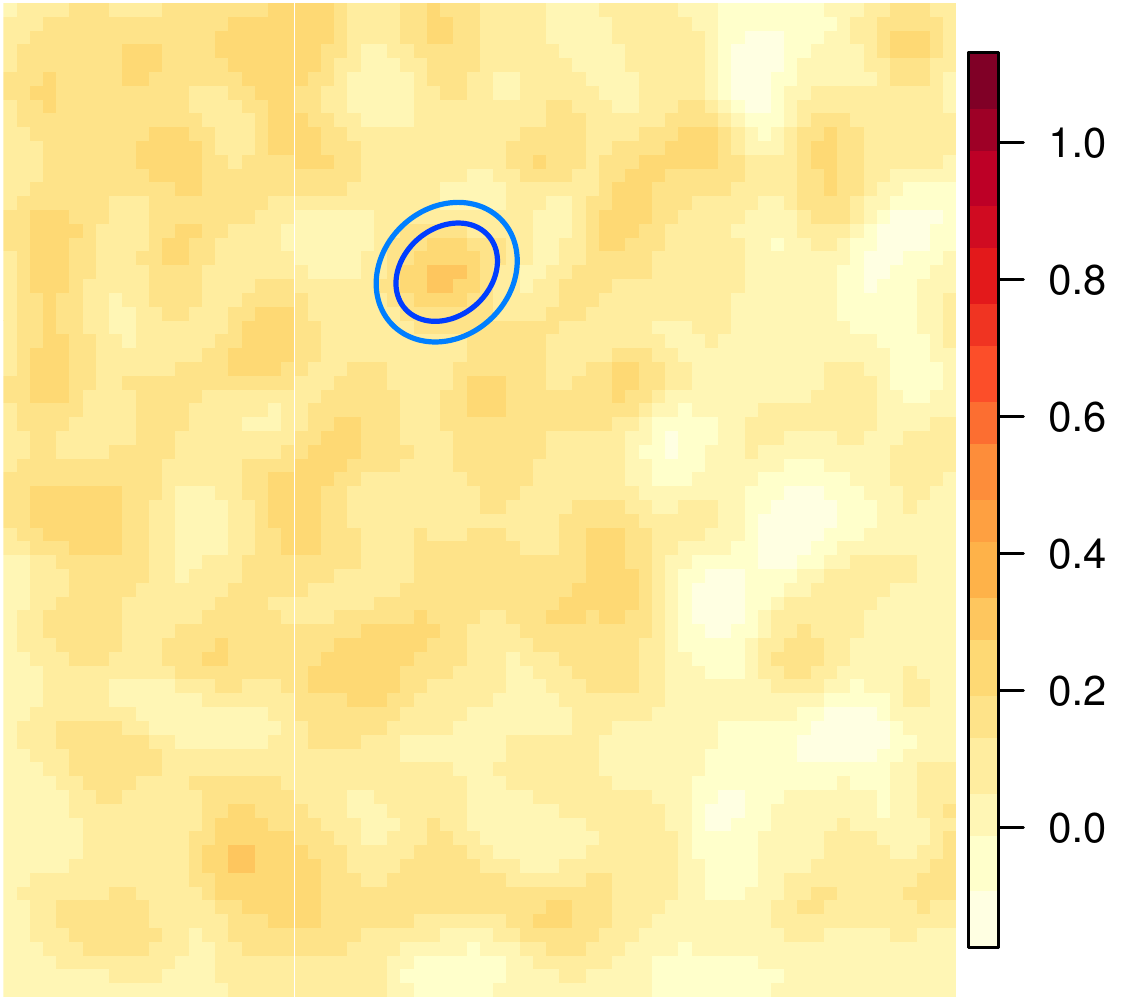}}}
\end{center}
\vspace{-0.2in}
\caption{The effect of $\lambda$ on optimal ellipses drawn on the 
  illustrative images of Figures~\ref{fig1} and~\ref{fig2} having
  strong (top row, a--c), weak (middle row, d--f) and no true (bottom row,
  g--i) signal.} 
\label{lambdaeffect}
\vspace{-0.1in}
\end{figure}
The objective function~\eqref{voleqn} in the main paper depends on
$\lambda$ so we now illustrate the effect of $\lambda$ with a bid to
make recommendations for its selection. 
Figure~\ref{lambdaeffect}  displays  the results -- for three different
values of $\lambda$ ($\lambda = 2$, first column; $\lambda = 100$,
second column; $\lambda = 200$, third column) -- of drawing the optimal
inner ellipse and the 
corresponding outer ellipse (using Step~\ref{step3} of our algorithm)
after optimizing $\mV(a,b,\theta)$ for the three illustrative cases of 
Section~\ref{vbexample} after processing with a matched filter,
resulting in images as in Figure~\ref{fig2}.  For presentation
clarity, the images in the top row with the strong true signal is drawn
using a different scale than the other two sets of images. 
The figures in the first column all have larger ellipses than their 
corresponding counterparts in the next two columns which have ellipses
in decreasing order of size. Thus, a larger regularization parameter results  
in smaller ellipses, because of the heavier penalty put on noise
pixels (which have negative $C_\sigma(\tau(u,v))$ values). The
differences in the sizes of the inner ellipses, however, decrease 
for larger values of $\lambda$ (last two columns). 

\subsection{Diagnostic Checks for Model Assumptions in
  Vibrothermography Specimens}
\label{supp.diag}
We report results on some checks to evaluate the assumption of normality
\begin{figure}[h]
\begin{center}
\mbox{
 \subfigure[]{\includegraphics*[width=0.5\textwidth]{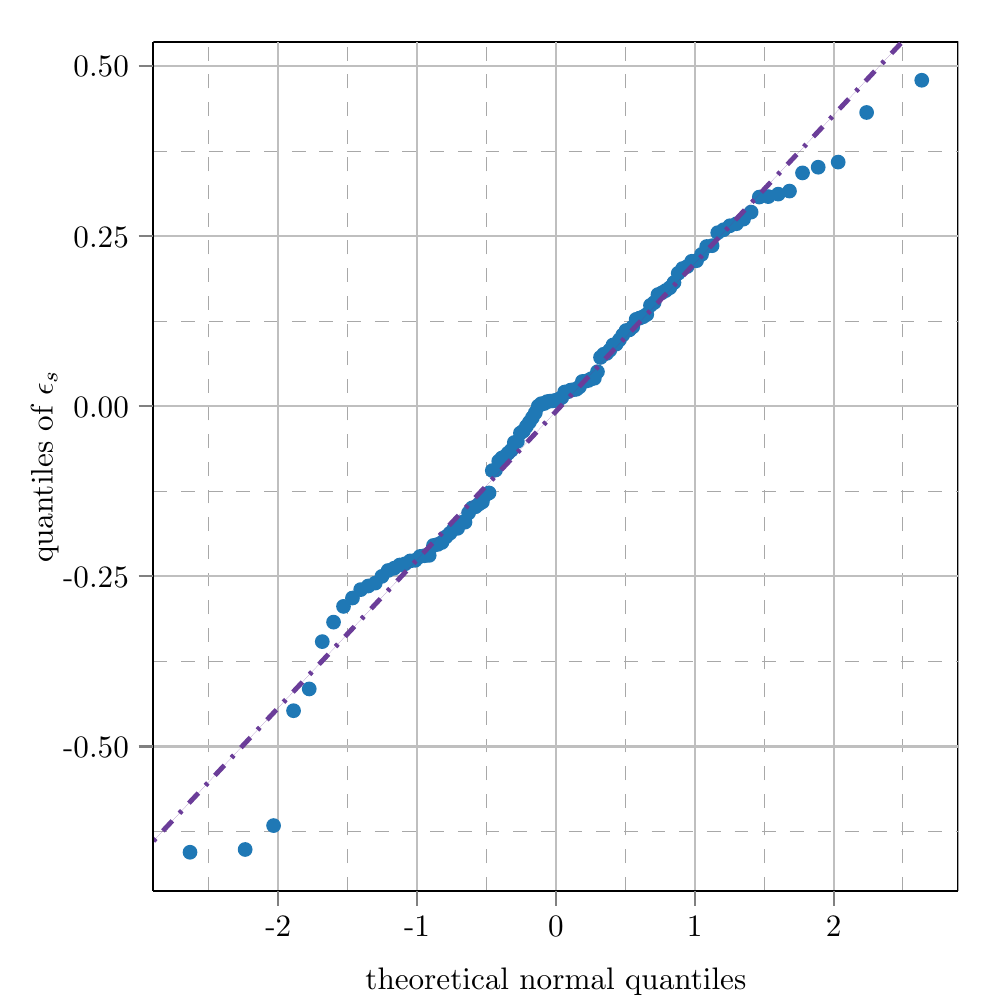}}
  \subfigure[]{\includegraphics*[width=0.5\textwidth]{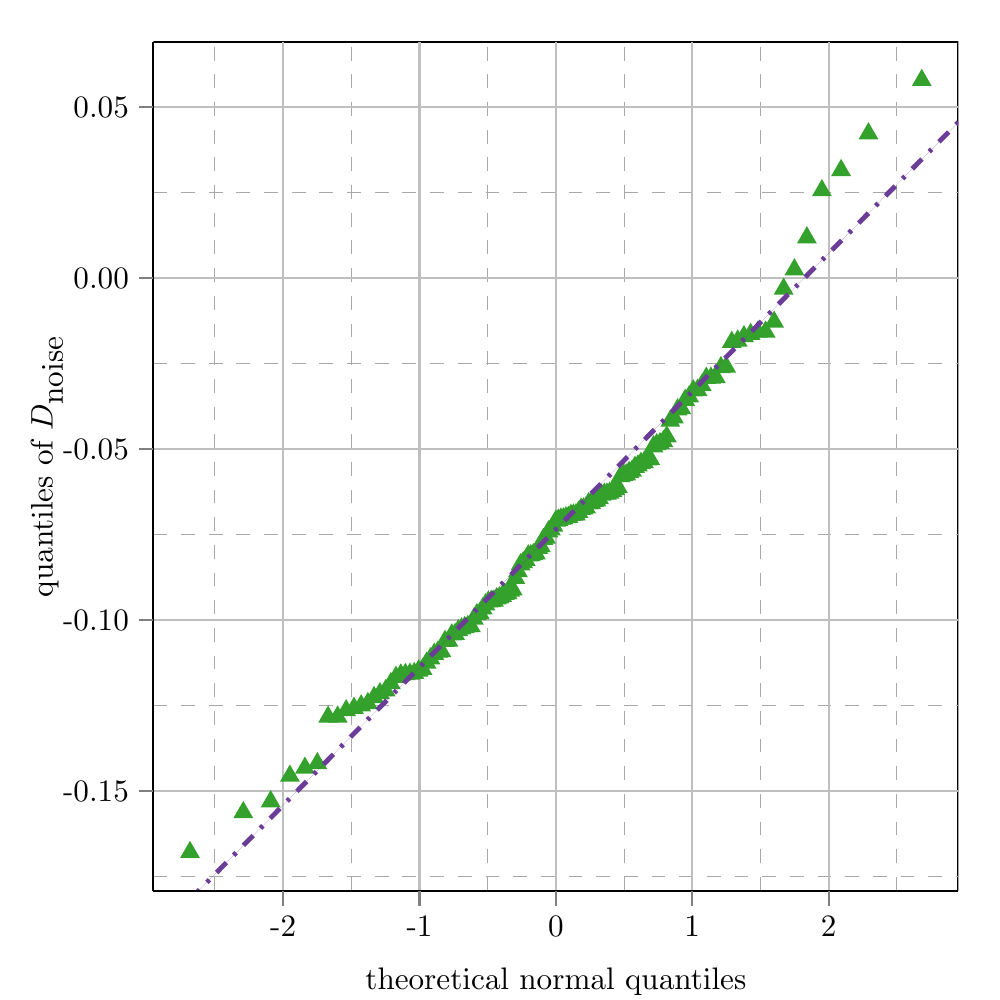}}}
\end{center}
\vspace{-0.2in}
\caption{Quantile plots for evaluating the assumption of normality in the NIM
  model for the (a) flawed  and (b) flawless specimens.}
\label{diags}
\vspace{-0.1in}
\end{figure}
in the NIM model for the vibrothermography specimen images. Figure~\ref{diags} provides quantile-quantile plots
for the residuals obtained upon fitting the NIM to  flawed and
flawless specimens. (Note that since there is no signal
in a flawless specimen, the residuals are essentially the same as
$D_{\mbox{noise}}$.) These plots indicate reasonably good agreement  
with the normal distribution. A formal test for normality using the
\citet{shapiroandwilk65} approach yielded $p$-values of 0.109  and
0.07 for flawed and flawless specimens, providing support for the
normality assumption. (We recall that in most real applications, the
Shapiro-Wilk test  will tend to find departures from a normal
distribution with moderately large to large sample sizes.) We conclude
by noting that we are not making inferences on the tails of the
distribution and so the procedure is expected to be robust to moderate
departures from the  normality assumption in the NIM model.

\section{Performance Evaluations -- Supplement}
\subsection{Derivation of the NIM Model}
\label{supp.deriv}
 We have, from the NIM model in Section~\ref{NIM}
 \begin{equation*}
  D_{\mbox{obs}} =
  \max(D_{\mbox{signal}},D_{\mbox{noise}})
  \end{equation*}
  where
\begin{equation*}
  D_{\mbox{signal}} =
  \beta_0+\beta_1 \log_{10}(\mbox{(flaw size)}) + \epsilon_s,
\end{equation*}
with $\epsilon_s\sim N(0,\sigma_s^2)$ distributed independently of
$D_{\mbox{noise}} \sim N(0,\sigma_N^2)$. Then the POD is given by,
\begin{equation*}
\begin{split}
  \mbox{POD(flaw)} & = \Pr(D_{\mbox{obs}} > 0)\\
  & = \Pr(\max(D_{\mbox{signal}},D_{\mbox{noise}}) > 0) \\
  & = 1-\Pr(\max(D_{\mbox{signal}},D_{\mbox{noise}}) \leq 0)\\
  & =  1-{\rm \Phi }\left[ 
-\frac{\beta _{0} +\beta _{1} \log_{10} ({\rm flaw})}{\sigma _{S} 
} \right]\Phi \left(-\frac{\mu _{N} }{\sigma _{N} } \right). \\
\end{split}
\end{equation*}
 A similar argument holds for the POD(flaw) of the ``PeakAmp''
 method, which is derived in Equation (3) of \cite{liandmeeker09}. To
 elucidate, since the 
 peak intensity $\check Z$ of the raw unprocessed hottest 
 frame image  (in $\log_{10}$) is used to characterize the image, we
 have that a flaw is detected if $\log_{10}\check Z_{\mbox{obs}}$ is
 greater than some threshold given by $\log Z_{\mbox{th}}$. Then, from
 the NIM, we have $\log_{10}\check Z_{\mbox{obs}} = 
 \max(Y_{\mbox{signal}},Y_{\mbox{noise}})$ so that 
  \begin{equation*}
\begin{split}
  \mbox{POD(flaw)} & = \Pr[Z_{\mbox{obs}} > Z_{\rm th}]\\
  & = \Pr[\log_{10}(Z_{\mbox{obs}}) > \log_{10}(Z_{\rm th})]\\
  & = \Pr[\max(Y_{\mbox{signal}},T_{\mbox{noise}}) > \log_{10}(Z_{\rm th})] \\
  & = 1-\Pr[\max(Y_{\mbox{signal}},Y_{\mbox{noise}} ) \leq \log_{10}(Z_{\rm th})]\\
  & =  1-{\rm \Phi }\left[ 
\frac{\log_{10}Z_{\rm th} - \gamma _{0} - \gamma _{1} \log_{10} ({\rm flaw})}{\kappa _{S} } \right]\Phi \left(\frac{\log_{10}Z_{\rm th} -\nu _{N} }{\kappa _{N} } \right). \\
\end{split}
\end{equation*}
where 
$(\gamma_0,\gamma_1,\kappa_S,\nu_N,\kappa_N)$ are defined as in the
last paragraph of~Section~\ref{performance}.
\renewcommand{\baselinestretch}{1}\normalsize
\begingroup
\def\bibfont{\small}
\bibliographystyle{asa}
\addcontentsline{toc}{section}{References}
\bibliography{refs}
\endgroup